\documentclass[a4paper,11pt]{article}
\pdfoutput=1
\usepackage{graphicx, rotating,amssymb,amsmath}

\long\def\symbolfootnote[#1]#2{\begingroup\def\thefootnote{\fnsymbol{footnote}}\footnote[#1]{#2}\endgroup}

\ifx\pdfoutput\undefined
\usepackage[dvips,bookmarks]{hyperref}	
\else
\usepackage{hyperref}	
\fi
\hypersetup{colorlinks,bookmarksopen,bookmarksnumbered,citecolor=red,linkcolor=verdes,pdfstartview=FitH,urlcolor=rosso}

\def\hhref#1{\href{http://arxiv.org/abs/#1}{#1}} 
\def\mhref#1{\href{mailto:#1}{#1}}		

\graphicspath{{/Users/p_salati/Dropbox/antiprotons_gaelle_mathieu/article/},{figures/},{./}}
\usepackage{multicol}
\usepackage{color}
\definecolor{rosso}{cmyk}{0,1,1,0.4}
\definecolor{rossos}{cmyk}{0,1,1,0.55}
\definecolor{rossoc}{cmyk}{0,1,1,0.2}
\definecolor{blu}{cmyk}{1,1,0,0.3}
\definecolor{blus}{cmyk}{1,1,0,0.6}
\definecolor{bluc}{cmyk}{1,0.4,0.1,0.1}
\definecolor{verde}{cmyk}{0.92,0,0.59,0.25}
\definecolor{verdec}{cmyk}{0.92,0,0.59,0.15}
\definecolor{verdes}{cmyk}{0.92,0,0.59,0.4}
\newcommand{\beq}{\begin{equation}}
\newcommand{\eeq}{\end{equation}}
\newcommand{\etal}{{\it et al.}}
\newcommand{\ie}{i.e.}

\newcommand{\GeV}{\,{\rm GeV}}

\newcommand{\pbar}{\bar{p}}
\newcommand{\nbar}{\bar{n}}

\def\PAMELA{{\sc Pamela}}
\def\AMS{{\sc Ams-02}}

\begin{document}
\begin{flushright}
\footnotesize
SACLAY--T14/099\\
LAPTh--236/14
\end{flushright}
\color{black}
\vspace{0.3cm}

\begin{center}
{\huge\bf A fussy revisitation of antiprotons\\[3mm] as a tool for Dark Matter searches}

\medskip
\bigskip\color{black}\vspace{0.6cm}

{
{\large\bf Mathieu Boudaud}\ $^{a}\, \symbolfootnote[1]{\mhref{mathieu.boudaud@lapth.cnrs.fr}}$,
{\large\bf Marco Cirelli}\ $^{b}\, \symbolfootnote[2]{\mhref{marco.cirelli@cea.fr}}$,
{\large\bf Ga\"elle Giesen}\ $^{b}\, \symbolfootnote[3]{\mhref{gaelle.giesen@cea.fr}}$,
{\large\bf Pierre Salati}\ $^{a}\, \symbolfootnote[4]{\mhref{pierre.salati@lapth.cnrs.fr}}$
}
\\[7mm]
{\it $^a$ \href{https://lapth.cnrs.fr}{LAPTh}, Universit\'e  Savoie Mont Blanc, CNRS,\\ BP 110, 74941 Annecy-le-Vieux, France}\\[3mm]
{\it $^b$ \href{http://ipht.cea.fr/en/index.php}{Institut de Physique Th\'eorique}, CNRS, URA 2306 \& CEA/Saclay,\\ 
F-91191 Gif-sur-Yvette, France}\\[3mm]

\end{center}

\bigskip
 
\centerline{\large\bf Abstract}
\begin{quote}
\color{black}\large
Antiprotons are regarded as a powerful probe for Dark Matter (DM) indirect detection and indeed current data from \PAMELA\ have been shown to lead to stringent constraints. However, in order to exploit their constraining/discovery power properly and especially in anticipation of the exquisite accuracy of upcoming data from \AMS, great attention must be put into effects (linked to their propagation in the Galaxy) which may be perceived as subleading but actually prove to be quite relevant. 
Using a semi-analytic code for rapidity, we revisit the computation of the astrophysical background and of the DM antiproton fluxes. Like in the fully numerical standard calculations, we include the effects of: diffusive reacceleration, energy losses including tertiary component and solar modulation (in a force field approximation).
We show that their inclusion can somewhat modify the current bounds, even at large DM masses, and that a wrong interpretation of the data may arise if they are not taken into account.
%
%
At the present level of accuracy of the data from {\sc Pamela}, the inclusion of the above effects amounts to changing the constraints, with respect to the case in which they are neglected, of up to about 40\% at a DM mass of 1 TeV and 30\% at 10 TeV. When the {\sc Ams-02} level of precision is reached, including them would strengthen (lessen) the bounds on the annihilation cross section by up to a factor of 15 below (above) a DM mass of 300 GeV.
The numerical results for the astrophysical background are provided in terms of fit functions; the results for Dark Matter are incorporated in the new release of the \href{http://www.marcocirelli.net/PPPC4DMID.html}{{\sc Pppc4dmid}}.
\end{quote}

\newpage

\tableofcontents

\section{Introduction}
\label{sec:introduction}

\noindent
The evidence for Dark Matter (DM) comes nowadays from a number of different astrophysical and cosmological probes, sensitive
to its gravitational effects. However, we are still eagerly awaiting an explicit manifestation of it. A possibility would be detecting anomalous
fluxes of cosmic rays (charged antimatter, photons, neutrinos\ldots), which is the aim of the so-called Indirect Detection strategy. Such anomalous
fluxes could indeed originate from DM pair annihilations or decays in the Milky Way halo, subsequently propagated to us within the Galactic
environment.

In particular, antiprotons are a sensitive probe for DM. Indeed, since the initial proposal~\cite{pbar_history}, many studies have stressed the
importance of this channel, including several recent ones~\cite{pbar_history2,Cirelli:2013hv,Bringmann:2014lpa,hooperon,Hooper:2014ysa,Zeldovich-Khlopov,Cerdeno:2011tf,Delahaye:2013yqa}.  This is both for intrinsic and contingent reasons.
An intrinsic reason is that the production of antiprotons is rather universal in DM models: as long as DM particles annihilate or decay into quarks or gauge
bosons (but also into leptons, thanks to ElectroWeak corrections, i.e. the emission of EW gauge bosons from the final state particles), $\bar p$ copiously emerge from the hadronization process. Other reasons are that the determination of the astrophysical background is relatively under control (at least if compared to other channels) and that the Galactic propagation of
antiprotons can be better modeled than the one of other charged particles. We will actually come back to these last two points in great detail, as they
constitute one of the main focus of our paper. A contingent reason, on the other hand, is that in other channels (most notably positrons and gamma rays)
sizable excesses have shown up, which cannot be easily attributed neither to DM nor to known astrophysical processes. Until their origin is clarified, they
greatly limit the robustness of DM analyses based on these channels. Finally, another motivation stems from the great precision of the $\bar p$ data already
available from the \PAMELA\ satellite and the even better precision expected soon from \AMS.

In this context, it is clearly timely to refine previous predictions of antiproton production from astrophysics and from DM, in order to obtain fluxes as accurate
as possible to be compared with the precise data. This is what we aim to do in this work. In particular, we upgrade previous computations by incorporating
energy losses and diffusive reacceleration, which will be discussed in detail below. We anticipate that these effects have a sizable impact, especially on
the low energy portion of the spectrum. Hence, they cannot be neglected if one aims at precision predictions.
One should nevertheless keep in mind that other sources of uncertainties could also possibly play a role, yet to be determined,
such as the nuclear antiproton production cross sections or the way cosmic ray propagation is modeled.

\medskip
  
The rest of this paper is organized as follows. In sec.~\ref{sec:propagation} we discuss the main concepts related to the propagation of antiprotons in the Galactic
environment, in particular the phenomena of energy losses, diffusive reacceleration and, to some extent, solar modulation. In sec.~\ref{sec:antip_components}
we discuss the antiproton inputs, both from Dark Matter and from astrophysics. In sec.~\ref{sec:constraints} we derive updated constraints on Dark Matter (using
existing \PAMELA\ data) and updated expected sensitivities (using projected \AMS\ mock data). Finally, in sec.~\ref{sec:conclusions} we conclude.

We provide numerical results in the form of fit functions for the astrophysical fluxes (see sec.~\ref{sec:antip_background}) and in the form of a new release
of the {\sc Poor Particle Physicist Cookbook for Dark Matter Indirect Detection} (\href{http://www.marcocirelli.net/PPPC4DMID.html}{{\sc Pppc4dmid}})
for what concerns the DM fluxes.

\section{Antiproton propagation in the Galaxy}
\label{sec:propagation}

In this section we address the main points concerning how we treat the propagation of antiprotons, from the production points in the Galactic halo or disk
to detection near Earth. The production itself will be addressed in the next section.
We first review the basic formalism for cosmic ray (CR) propagation in the Galaxy (sec.~\ref{sec:transport}).
We then focus more specifically on the process of energy losses, tertiary production and diffusive reacceleration (sec.~\ref{sec:ELDR})
as well as solar modulation (sec.~\ref{sec:SolarMod}).

\subsection{Basics of cosmic ray transport in the Galaxy}
\label{sec:transport}

Antiprotons, like any other charged species, are deflected by the Galactic magnetic field and their transport may be seen as a diffusion process
where the irregularities of this turbulent field play the role of scattering centers. In full generality, the master equation for the energy and space
distribution function $f = dn/dK$ of any species of charged cosmic rays can be written as
\begin{equation}
\frac{\partial f}{\partial t} - \mathcal{K}(K) \! \cdot \! \nabla^2f +
\frac{\partial}{\partial z}\left\{ {\rm sign}(z) \, f \, V_{\rm conv} \right\} +
\frac{\partial}{\partial E}\left\{ b(K , \vec{x}) f - \mathcal{K}_{\rm EE}(K) \frac{\partial f}{\partial E} \right\} = Q,
\label{eq:transport}
\end{equation}
The first term of the left hand side is put to zero since one is interested in steady state conditions.
The second term accounts for space diffusion and, for antiprotons, can be simply modeled as $\mathcal{K}(K) = \mathcal{K}_0 \, \beta \, ({p}/{\rm GeV})^\delta$
in terms of the kinetic energy $K$, the beta factor and the momentum $p$ of the particle. The parameters $\mathcal{K}_0$ and $\delta$ set the normalization
and momentum dependence.
The third term corresponds to the convective processes, with characteristic velocity $V_{\rm conv}$, which originate in the disk and tend to push vertically
outwards (hence the $z$ gradient) the antiproton fluxes. The resulting Galactic wind reaches its nominal value of $\pm V_{\rm conv}$ right outside the disk,
assumed here to be infinitely thin.
The fourth term (inside the energy derivative) accounts for energy losses, which are in general energy and space dependent.
The thin disk approximation leads to write $b(K , \vec{x})$ as $2 h \delta(z) \, b(K)$, where $h = 100$~pc is the half-height of the disk. 
Notice that energy losses are typically not very important for antiprotons and are generally neglected. We will nevertheless fully include them in our detailed analyses.
%
The last term of the l.h.s. represents diffusive reacceleration and is discussed in detail below.

On the right hand side, the equation features the source term $Q$, which can contain different contributions.
The spallation of high-energy cosmic rays on the interstellar gas produces antiprotons (so called `secondary') which are the source of the astrophysical
background. The annihilations or decays of DM produce (so called `primary') antiprotons. These two components are discussed in sec.~\ref{sec:antip_components}.
$Q$ contains also a sink term, due to the annihilations of the antiprotons on the interstellar gas. Such a term reads $-2h \delta(z)\, \Gamma_{\rm ann}$ where
the $\delta$ function effectively localizes the interactions only in the disk where the ISM sits. The annihilation rate $\Gamma_{\rm ann}$ is equal to
$(n_{\rm H} + 4^{2/3} n_{\rm He}) \sigma^{\rm ann}_{\bar{p} p} v_{\bar{p}}$, where $n_{\rm H}$ = 0.9 cm$^{-3}$ and $n_{\rm He}$ = 0.1 cm$^{-3}$ stand for the ISM hydrogen and helium densities, 
while $v_{\bar{p}}$ denotes the velocity of the incoming antiproton. The annihilation cross section
$\sigma^{\rm ann}_{\bar{p} p}$ is borrowed from \cite{Tan_Ng_82,Tan_Ng_83}, and we have multiplied it by a factor of $4^{2/3} \sim 2.5$ to account
for the different geometrical cross section on helium in an effective way. Notice that antiprotons also collide elastically on interstellar H and He. Since they are
preferentially scattered forward, this process has no effect and does not contribute to the sink term in $Q$.
Last but not least, $Q$ contains a source term (or rather `recycling' term) corresponding to tertiary antiprotons, which will be discussed in the next subsection
together with energy losses and diffusive reacceleration.

In order to solve the transport equation~(\ref{eq:transport}), we model the magnetic halo of the Milky Way by a flat cylinder with
half-height $L$ and radius $R$ = 20 kpc, inside which cosmic rays diffuse. The Galactic disk lies in the middle at $z=0$ and is assumed to
be infinitely thin as discussed above. The CR densities $f \equiv {dn}/{dK}$ are assumed to be axi-symmetric. They are expanded
along the radial direction as a series of Bessel functions of zeroth-order
\beq
f(r , z , E) = {\displaystyle \sum_{i=1}^{+ \infty}} \; F_{i}(z , E) \, J_{0} \left( \alpha_{i} \, r / R \right).
\label{bessel_psi}
\eeq
Since $\alpha_{i}$ is the $i^{\mathrm{th}}$ zero of $J_{0}$, the density
vanishes at $r = R$. The Bessel transforms $F_{i}(z , E)$ also vanish at the vertical boundaries $z = \pm L$ of the diffusive halo.
The transport equation is solved for each Bessel order $i$ and the antiproton flux at the Earth is derived as explained
in~\cite{Donato:2001ms,Bringmann:2006im}.

%
\begin{table}[t]
\begin{center}
\begin{tabular}{c|ccccc}
 &  \multicolumn{5}{c}{Antiproton propagation parameters}  \\
Model  & $\delta$ & $\mathcal{K}_0$ [kpc$^2$/Myr] & $V_{\rm conv}$ [km/s] & $L$ [kpc]  & $v_{a}$ [km/s] \\
\hline 
MIN  &  0.85 &  0.0016 & 13.5 & 1 & 22.4 \\
MED &  0.70 &  0.0112 & 12 & 4  & 52.9 \\
MAX  &  0.46 &  0.0765 & 5 & 15 & 117.6
\end{tabular}
\caption{\em \small {\bfseries Cosmic ray transport parameters} for antiprotons in the Galactic halo (from~\cite{DonatoPRD69}).
Here $\delta$ and $\mathcal{K}_0$ are the power index and the normalization of the diffusion coefficient, $V_{\rm conv}$ is the velocity of the
convective wind, $L$ is the half-thickness of the diffusive cylinder and $v_a$ is the velocity of the reaccelerating scattering centers.
\label{tab:proparam}}
\end{center}
\end{table}
%

\subsection{The Energy Losses including tertiaries, and Diffusive Reacceleration (`ELDR')}
\label{sec:ELDR}

Three processes of energy loss are encoded in the negative coefficient $b$. First, like any other nuclear species, antiprotons undergo ionization losses in the interstellar neutral matter, whose composition has been given above.
Then, Coulomb energy losses take place on the fraction of the interstellar medium (ISM) that is completely ionized. That mechanism is dominated by scatterings
on thermal electrons, for which we have used a density of 0.033 cm$^{-3}$ and a temperature of $3 \times 10^{5}$~K.
These two effects are discussed in \cite{Mannheim:1994sv} and \cite{Strong:1998pw}, where complete expressions may be found for the energy loss rate $b$.
Finally, convective processes also induce a loss of energy through the conservation of the CR density in phase-space. This leads to
\beq
b_{\rm adia}(K , \vec{x}) = - \frac{1}{3} \, \left\{ \nabla_{x} \! \cdot \! \vec{u}(\vec{x}) \right\} \, \frac{p^{2}}{E} \stackrel{\rm ave}{\xrightarrow{\hspace*{1cm}}} b_{\rm adia}(K) = - \frac{V_{\rm conv}}{3h} \, \frac{p^{2}}{E}
\eeq
where $p$ and $E$ denote respectively the momentum and energy of the antiproton and the last expression is obtained once the divergence of the convective wind $\vec{u}$ is averaged across the thin disk.

As anticipated, the term `tertiary antiprotons' identifies the particles emerging from inelastic and non-annihilating interactions of primary or secondary
antiprotons on the ISM. An antiproton can collide on a proton at rest and transfer enough energy to excite it as a $\Delta$ resonance. The $\bar p$ typically
loses a fraction of its energy and is effectively reinjected in the flux with a degraded momentum. This mechanism redistributes antiprotons towards lower
energies, hence flattening their spectrum as first remarked by \cite{ber99}. The rate for the production of tertiary antiprotons is given by
\beq
Q_{\pbar}^{\rm ter}(\vec{x} , E_{\pbar}) = {\displaystyle \int_{E_{\pbar}}^{+ \infty}} \,
{\displaystyle \frac{d\sigma_{\bar{p} p \to \bar{p} X}}{dE_{\pbar}}}(E'_{\pbar} \! \to \! E_{\pbar}) \,
n_{\rm H}(\vec{x}) \, v'_{\pbar} \, f(\vec{x} , E'_{\pbar}) \, dE'_{\pbar} \; - \;
\sigma_{\bar{p} p \to \bar{p} X}(E_{\pbar}) \, n_{\rm H}(\vec{x}) \, v_{\pbar} \, f(\vec{x} , E_{\pbar}).
\label{Q_tertiary}
\eeq
In this expression, the differential cross section for inelastic and non-annihilating interactions has been approximated by
\beq
{\displaystyle \frac{d\sigma_{\bar{p} p \to \bar{p} X}}{dE_{\pbar}}}(E'_{\pbar} \! \to \! E_{\pbar}) =
{\displaystyle \frac{\sigma_{\bar{p} p \to \bar{p} X}(E'_{\pbar})}{K'_{\pbar}}}.
\eeq
Following~\cite{ber99}, we assume that an antiproton undergoing such a reaction has final kinetic energy $K_{\pbar}$ uniformly distributed between 0
and its initial value $K'_{\pbar}$. The total inelastic and non-annihilating cross section $\sigma_{\bar{p} p \to \bar{p} X}$ has been borrowed from
\cite{Tan_Ng_83} where, above a kinetic energy of 13.3~GeV, it is approximated by the inelastic proton-proton cross section.
%
%
At high energy, the second term in relation~(\ref{Q_tertiary}) is dominant. The antiproton energy distribution is depleted
for the benefit of the low-energy tail of the spectrum where the first term contributes most.
Notice also that the tertiary production rate is proportional to the hydrogen density, which should be rescaled by a mere factor of
$(n_{\rm H} + 4^{2/3} n_{\rm He})$ to take also into account the helium component of the ISM. Finally, a global factor of
$2h \delta(z)$ should be added in the thin disk approximation.

As mentioned in sec.~\ref{sec:transport}, the last term in the l.h.s. of the transport equation~(\ref{eq:transport}) accounts
for diffusive reacceleration. This mechanism is produced by the drift with velocity $v_{a}$ of the diffusion centers, {\ie} the knots
of the turbulent Galactic magnetic field. This yields a second order Fermi acceleration which boils down into a diffusion in energy space
whose coefficient may be expressed as
\beq
\mathcal{K}_{\rm EE}(K) = \frac{2}{9} \, v_{a}^{2} \, {\displaystyle \frac{E^{2} \beta^{4}}{\mathcal{K}(K)}}.
\label{eq:K_EE}
\eeq
The antiproton energy and velocity are respectively denoted by $E$ and $\beta$, whereas the space diffusion coefficient $\mathcal{K}$
appears at the denominator (essentially expressing the fact that the more efficiently cosmic rays diffuse in space, the fewer collisions there are and the weaker the energy
diffusion). Other forms are possible and may be found for instance in~\cite{Strong:1998pw,PDM_MCMC_1}.
They are inspired by the diffusion coefficient $\mathcal{K}_{\rm pp}$ derived in momentum space by~\cite{seo_ptuskin}. We will
nevertheless keep relation~(\ref{eq:K_EE}) for our study because it is this particular form that has been used in~\cite{Maurin:2001sj}
to constrain the CR propagation parameters from the B/C tracer, and to define in~\cite{DonatoPRD69} the canonical models MIN, MED
and MAX.
%
\begin{figure}[t]
\begin{center}
\includegraphics[width=0.6 \textwidth]{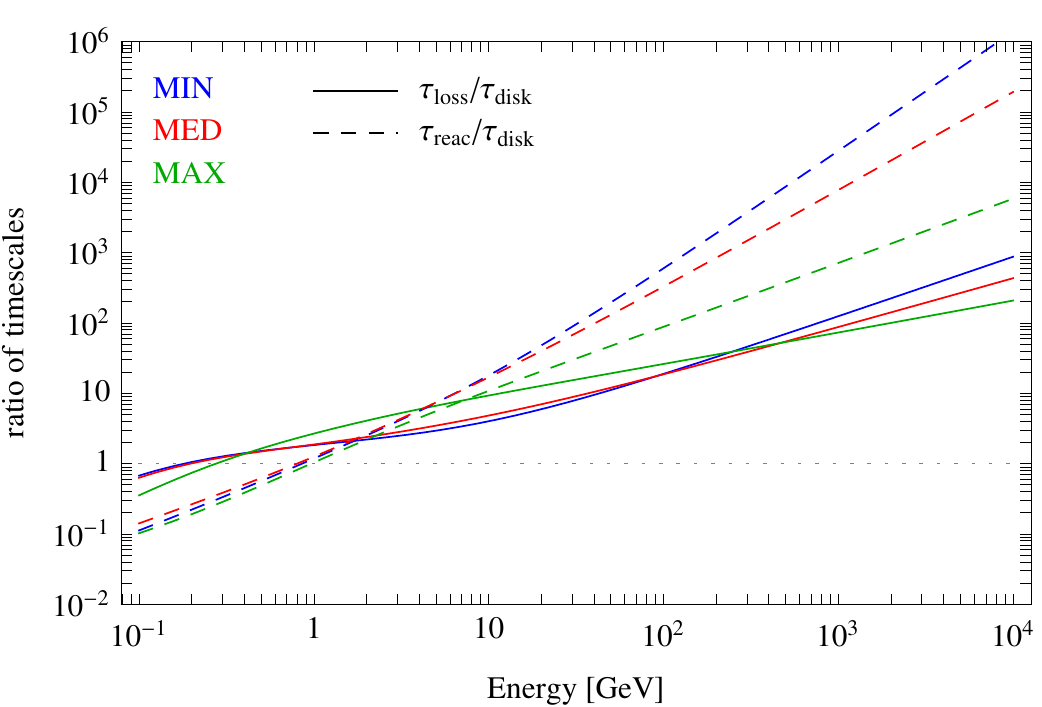} 
\caption{\small \em\label{fig:timescale}
The {\bfseries timescale ratios} ${\tau_{\rm \, loss}}/{\tau_{\rm disk}}$ (solid) and ${\tau_{\rm reac}}/{\tau_{\rm disk}}$ (short dashed)
are plotted as a function of the antiproton kinetic energy $K$. The blue, red and green curves respectively stand for the three
canonical models MIN, MED and MAX of cosmic ray propagation as defined in~\cite{DonatoPRD69}.
}
\end{center}
\end{figure}
%

As inelastic and non-annihilating interactions result into a discontinous variation of the antiproton energy, they will be associated
in this analysis to the continuous energy losses described by the $b(K)$ coefficient. The latter comes into play in the transport
equation together with diffusive reacceleration. These three processes will hence be switched on or off together, and are hereafter
denoted generically by the acronym `ELDR'.
In order to get a flavor of the relative importance of the various mechanisms which come into play in cosmic ray transport, we should
compare their timescales. To this end, we define the three timescales $\tau_{\rm disk}$, $\tau_{\rm loss}$ and $\tau_{\rm reac}$. 
Space diffusion and Galactic convection can be combined together in a one dimensional slab model
to yield the typical confinement timescale inside the disk
\beq
\tau_{\rm disk} = {\displaystyle \frac{h}{V_{\rm conv}}}
\left\{ 1 - \exp(- {\displaystyle \frac{V_{\rm conv} L}{\mathcal{K}}}) \right\}.
\eeq
At low energy, convection dominates and $\tau_{\rm disk} \simeq {h}/{V_{\rm conv}}$ is constant. Above a few GeV, diffusion
takes over since $\mathcal{K}$ is an increasing function of kinetic energy $K$, and the usual escape time
$\tau_{\rm disk} \simeq {h L}/{\mathcal{K}}$ of the Leaky Box model is recovered.
The typical timescale for energy losses is instead defined by $\tau_{\rm \, loss} = - {K}/{b(K)}$. Above a few GeV, ionization and Coulomb losses
are negligible with respect to adiabatic losses, and $\tau_{\rm \, loss}$ becomes equal to the constant ${3 h}/{V_{\rm conv}}$.
Finally, the diffusive reacceleration timescale $\tau_{\rm reac}$ is defined by the ratio ${K^{2}}/{\mathcal{K}_{\rm EE}(K)}$.
As soon as $\beta \simeq 1$, this timescale increases with energy like ${9 \mathcal{K}(K)}/{2 v_{a}^{2}}$.

The ratios ${\tau_{\rm \, loss}}/{\tau_{\rm disk}}$ (solid) and ${\tau_{\rm reac}}/{\tau_{\rm disk}}$ (short dashed) are displayed
in fig.~\ref{fig:timescale} as a function of antiproton kinetic energy $K$, for the three cosmic ray propagation canonical models MIN,
MED and MAX defined in~\cite{DonatoPRD69}.
Above 1~GeV, all ratios exceed unity, and $\tau_{\rm disk}$ is the smallest timescale. Convection and space diffusion are the dominant
processes, with the latter taking over the former above a few GeV. That is why simple approaches like the Leaky Box or infinite slab models
reproduce well the B/C observations. For kinetic energies smaller than 1~GeV,  the short dashed curves are below the dotted horizontal line
(with critical value 1) and diffusive reacceleration becomes the dominant mechanism.
Although never dominant, energy losses slightly deplete the high energy part of the antiproton spectrum and replenish its low energy tail,
flattening it. Possible bumps showing up in the injected spectra are partially erased once energy losses come into play. As expected, we also find
that inelastic but non-annihilating interactions, which are responsible for the tertiary production of antiprotons, have qualitatively the same effect,
albeit with a somewhat lesser extent.
Diffusive reacceleration tends to smooth globally the antiproton spectrum. Although inducing qualitatively the same effect at low energy
as the two other mechanisms, that process tends on the contrary to slightly replenish the high-energy part of the antiproton spectrum. As
featured by the ${\tau_{\rm reac}}/{\tau_{\rm disk}}$ curves of fig.~\ref{fig:timescale}, diffusive reacceleration is more and more effective
along the MIN--MED--MAX sequence of cosmic ray transport models.
Once all these mechanisms are combined, their global effect depends on the relative importance of diffusive reacceleration with respect
to energy losses. As indicated in fig.~\ref{fig:timescale}, the ${\tau_{\rm reac}}/{\tau_{\rm \, loss}}$ ratio significantly decreases from the
MIN to the MAX configurations. In the MIN and MED cases, the antiproton spectrum is slightly depleted at high energies when the `ELDR'
effects are included, whereas the opposite effect is observed for the MAX model where diffusive reacceleration counterbalances energy losses.

\subsection{Solar modulation (`SMod') effects in a force-field approximation}
\label{sec:SolarMod}

In the final portion of their journey, antiprotons penetrate into the sphere of influence of the Sun and are subject to the phenomenon of solar modulation
(denoted with the abbreviation `SMod' hereafter). In general terms, the solar CR wind and the solar magnetic field have the effect of decreasing the kinetic
energy and momentum of the particles, especially low energy ($\lesssim$ 10 GeV) ones.
This can be effectively described in the so-called `force field approximation'~\cite{GA}: the energy spectra in the local interstellar environment $d\Phi_{\rm LIS}/dK$
(i.e. at the end of the galactic propagation but before entering into the solar sphere) are modulated to obtain the flux at Earth $d\Phi_\oplus/dK$ in the following
way
\begin{equation}
\frac{d\Phi_\oplus}{dK}(K)=\frac{d\Phi_{\rm LIS}}{dK}(K+ |e| \phi_F Z/A)\cdot \frac{ K(K+2m)}{(K+m+ |e| \phi_F Z/A)^2-m^2}.
\label{eq:SMod}
\end{equation}
%
%
Here $Z$, $A$, $e$ and $m$ are the atomic number, the mass number, the electron charge and the mass of the CR species. In our case, $Z=A=1$ for protons and antiprotons.
The force-field or Fisk potential $\phi_F$ parameterizes the effect of the solar modulation on CRs and will take a value which depends on several
complex parameters of the solar activity and therefore ultimately on the epoch of observation~\footnote{Note that, with the notation $\phi_F$, we
will always refer in this work to the Fisk potential {\it for antiprotons}. The corresponding quantity for protons, $\phi_F^p$ can in principle be different
(in which case one has `charge dependent' solar modulation). Ref.~\cite{hooperon}, based on the same {\tt HelioProp} runs mentioned below, finds that
the two quantities typically do not differ by more than 50\%. Moreover, dedicated runs find that the value for antiprotons tends to be larger than the one for protons, at least for conditions of solar activity featuring a negative polarity of the solar magnetic field, a `tilt angle' of the heliospheric current sheet of about 20-40 degrees (both assumptions being appropriate for the \PAMELA\ data taking period) and a `parallel mean free path' (mfp) of protons at Earth not smaller than $\sim 0.05$ AU~\cite{Gaggero}. Only if the mfp assumes very small values one can have a Fisk potential for protons larger than the one for antiprotons. These details do not impact our subsequent analysis since we choose very conservative intervals for $\phi_F$ anyway.}.
For the analysis of the \PAMELA\ data, we choose a conservative interval $0.1\ {\rm GV} < \phi_F < 1.1\ {\rm GV}$. This is based on the fact that, using
more refined tools such as {\tt HelioProp}~\cite{Maccione:2012cu} to model the propagation in the heliosphere, ref.~\cite{hooperon} has concluded that
for the \PAMELA\ data taking period, the solar modulation conditions correspond to such a range. However, for the \AMS\ data, we will want to be even
more conservative and choose $0\ {\rm GV} <\phi_F<2$ GV.

\section{The antiproton astrophysical and DM components}
\label{sec:antip_components}

In this section we briefly review the computation of the two main antiproton input components: the background, from astrophysics, and the primary signal,
from DM annihilations or decays.

\subsection{The antiproton astrophysical flux}
\label{sec:antip_background}

The astrophysical antiproton background is produced by the collisions of high-energy CR protons and helium nuclei on the ISM,
which is assumed here to be mostly composed of hydrogen and helium. In the case of the interactions between CR protons and
hydrogen atoms, the source term takes the following form
\beq
Q_{\pbar}^{\rm sec}(\vec{x} , E_{\pbar}) = {\displaystyle \int_{E^{0}_{p}}^{+ \infty}} \,
{\displaystyle \frac{d\sigma_{p {\rm H} \to {\pbar} X}}{dE_{\pbar}}}(E_{p} \! \to \! E_{\pbar}) \, n_{\rm H}(\vec{x}) \,
v_{p} \, f_{p}(\vec{x} , E_{p}) \, dE_{p}.
\label{Q_secondary}
\eeq
We use the injection proton and helium CR fluxes at the Earth as measured by the \PAMELA\ experiment~\cite{pamela_p_he_2011}\footnote{Technically, we employ a numerical fit of those data performed by T.~Delahaye~\cite{Timur}.}. Following previous studies~\cite{Donato:2001ms,Bringmann:2006im}, the Bessel
transforms of these fluxes are calculated for each CR propagation model in order to derive the proton and helium densities $f_{p}$ and
$f_{\alpha}$ all over the Galactic magnetic halo. The radial profile of the sources of primary CR nuclei comes into play and can be
determined from pulsar and surpernova remnant surveys. We have used here the parameterization of~\cite{yusifov_kucuk}, slightly
modified by~\cite{Bernard:2012wt}.

In the case of a proton impinging on a hydrogen atom at rest, the production rate peaks around a few GeV. The energy of the projectile must
actually exceed a threshold of $E^{0}_{p} = 7 m_{p}$.
In the Galactic frame, where the target is at rest, the differential production cross section of the previous relation is given by the integral
\beq
{\displaystyle \frac{d \sigma_{p {\rm H} \to {\pbar} X}}{dE_{\pbar}}} = 2 \pi k_{\pbar} \,
{\displaystyle \int_{0}^{\theta_{\rm max}}}
\left( E_{\pbar} \, {\displaystyle \frac{d^{3} \sigma}{d^{3} {k}_{\pbar}}} \right)_{\rm LI}
d(- \cos \theta),
\label{dif_cross_section}
\eeq
where $\theta$ is the angle between the momenta of the incoming proton and the produced antiproton. In the center of mass frame,
which drifts with a velocity $\beta_{\rm CM} = \{ (E_{p} - m_{p}) / (E_{p} + m_{p}) \}^{1/2}$ with respect to the Galactic frame,
the antiproton energy cannot exceed a value of
\beq
E_{\pbar , {\rm max}}^{*} = {\displaystyle \frac{s \, - \, 9 \, m_{p}^{2} \, + \, m_{p}^{2}}{2 \sqrt{s}}},
\eeq
where $\sqrt{s} = \{ 2 m_{p} (E_{p} + m_{p}) \}^{1/2}$ is the total energy of the reaction. In eq.~(\ref{dif_cross_section}), the
energies $E_{p}$ and $E_{\pbar}$ have been fixed and the angular integral runs from $\theta = 0$ up to a maximal value of
$\theta_{\rm max}$ for which
\beq
\cos \theta_{\rm max} = {\displaystyle \frac{1}{\beta_{\rm CM} k_{\pbar}}} \,
\left( E_{\pbar} - {\displaystyle \frac{E_{\pbar , {\rm max}}^{*}}{\gamma_{\rm CM}}} \right).
\eeq
The Lorentz invariant differential cross section $E_{\pbar} \, ({d^{3} \sigma}/{d^{3} {k}_{\pbar}})$ depends on the antiproton rapidity
$y = \tanh^{-1}(k_{\pbar \, \parallel}/E_{\pbar})$ and transverse mass $m_{T}^{2} =  m_{p}^{2} + k_{\pbar \, \perp}^{2}$. We have used
a new parameterization recently proposed by~\cite{diMauro:2014zea} instead of the Tan and Ng fitting relations~\cite{Tan_Ng_82,Tan_Ng_83}.
As mentioned by~\cite{Bringmann:2006im}, the transverse mass $m_{T}$ should be preferred to the angular variable $\cos \theta$ in
the integral~(\ref{dif_cross_section}) whenever the maximal angle $\theta_{\rm max}$ is small, under penalty of numerical errors.
%
Antiprotons can also be produced in reactions involving helium nuclei either in the cosmic radiation or in the ISM. We have used the same
procedure as discussed in~\cite{Bringmann:2006im} to which we refer the reader for details.

Notice finally that antineutrons are also produced, and should be taken into account as they subsequently decay into antiprotons. It has been
so far conventionally assumed that the antineutron and antiproton production rates are equal insofar as isospin symmetry should hold. A global
factor of 2 was generally assumed in order to account for antineutrons. But measurements by the NA49 experiment~\cite{Fischer_2003} of the
differential antiproton multiplicity in $pp$ and $np$ collisions point towards a different conclusion, as noticed by~\cite{Kappl:2014hha}. In the
case of antineutrons, the production cross section should be multiplied by an unknown factor $N_{\rm IS}$ such that
\beq
{\displaystyle \frac{d \sigma_{p {\rm H} \to {\nbar} X}}{dE_{\nbar}}} = N_{\rm IS} \,
{\displaystyle \frac{d \sigma_{p {\rm H} \to {\pbar} X}}{dE_{\pbar}}}.
\eeq
For conservativeness, we have assumed $N_{\rm IS}$ to lie in the range from 1 to 1.5, so that our calculations can be rescaled by a factor
$A \equiv (1 + N_{\rm IS})/2$ which has been varied freely from 1 to 1.25 in order to improve the quality ot the fits.

\begin{figure}[t!]
\begin{center}
\includegraphics[width=0.32\textwidth]{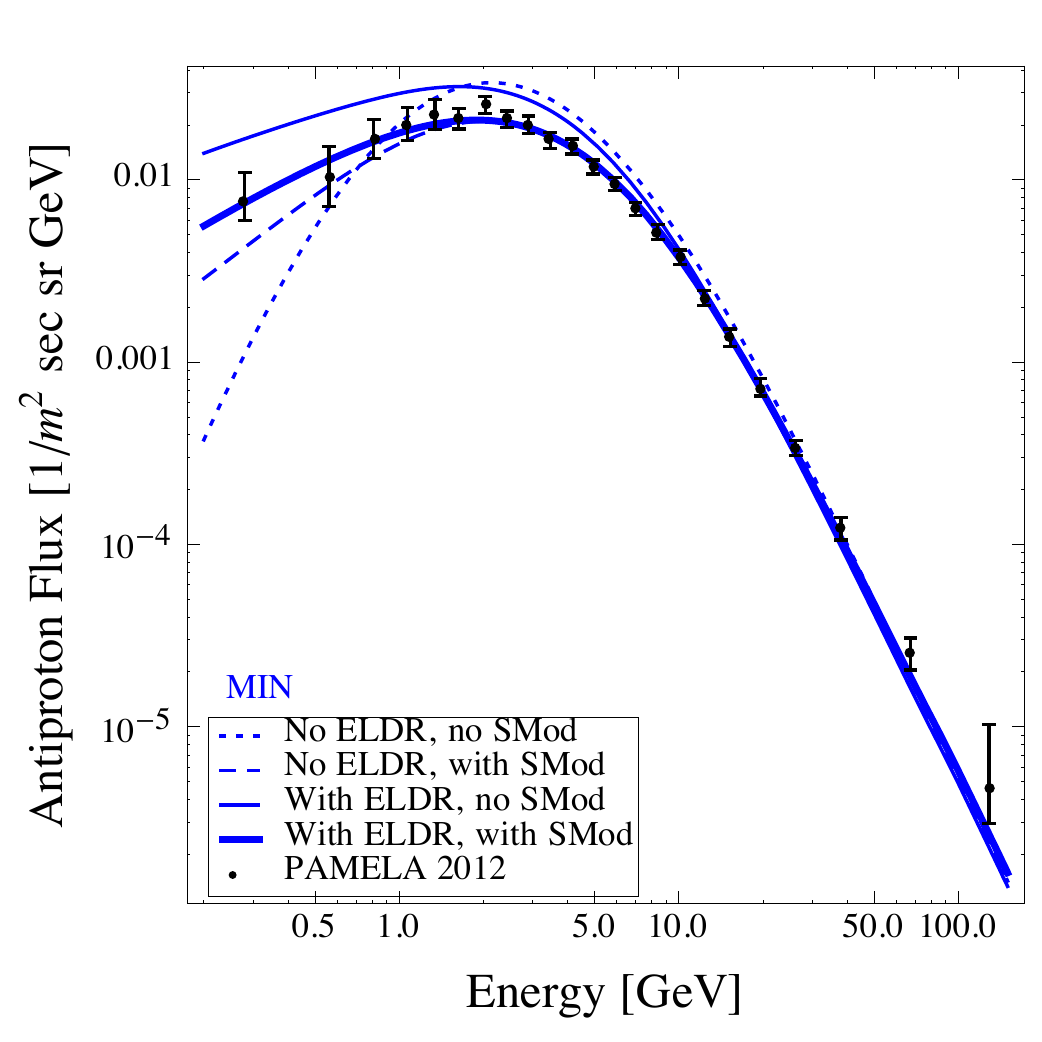} 
\includegraphics[width=0.32\textwidth]{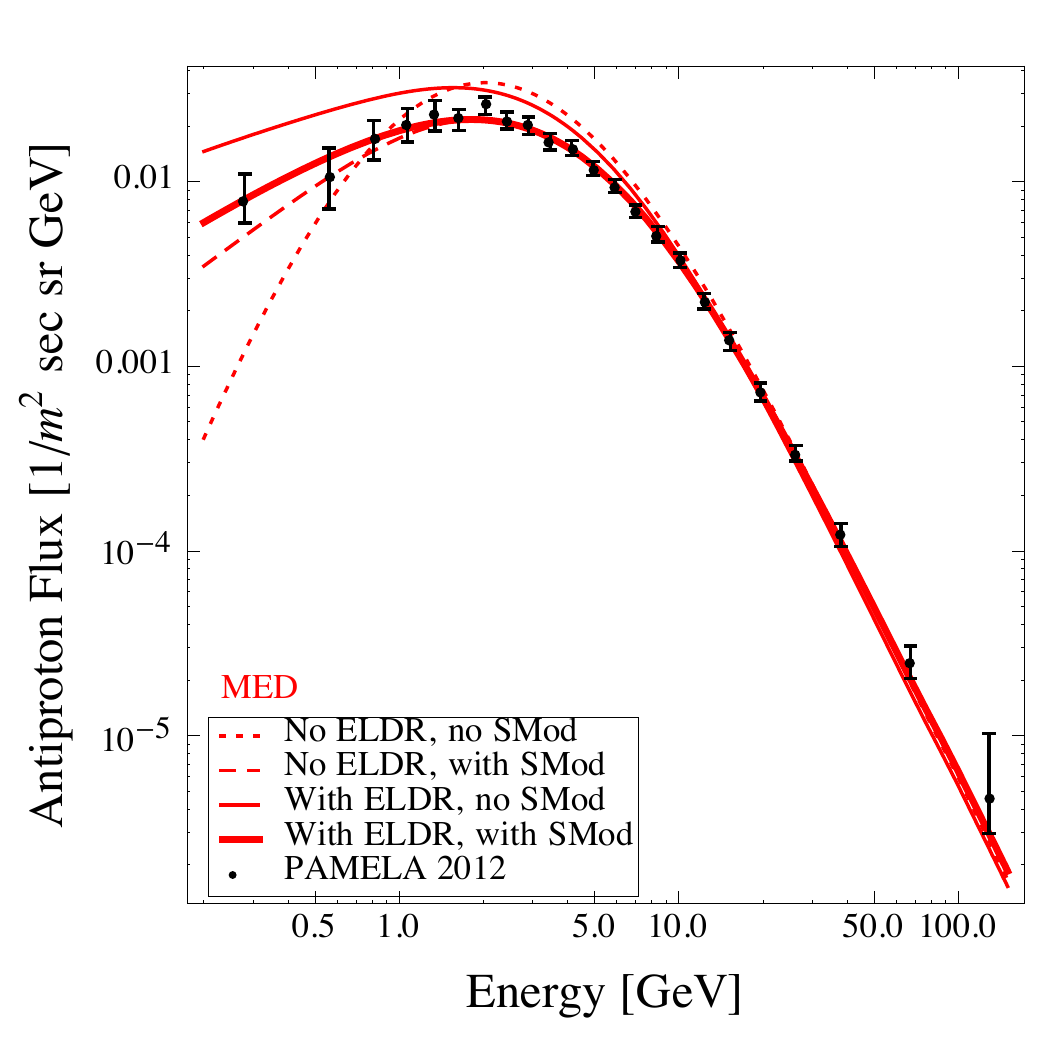} 
\includegraphics[width=0.32\textwidth]{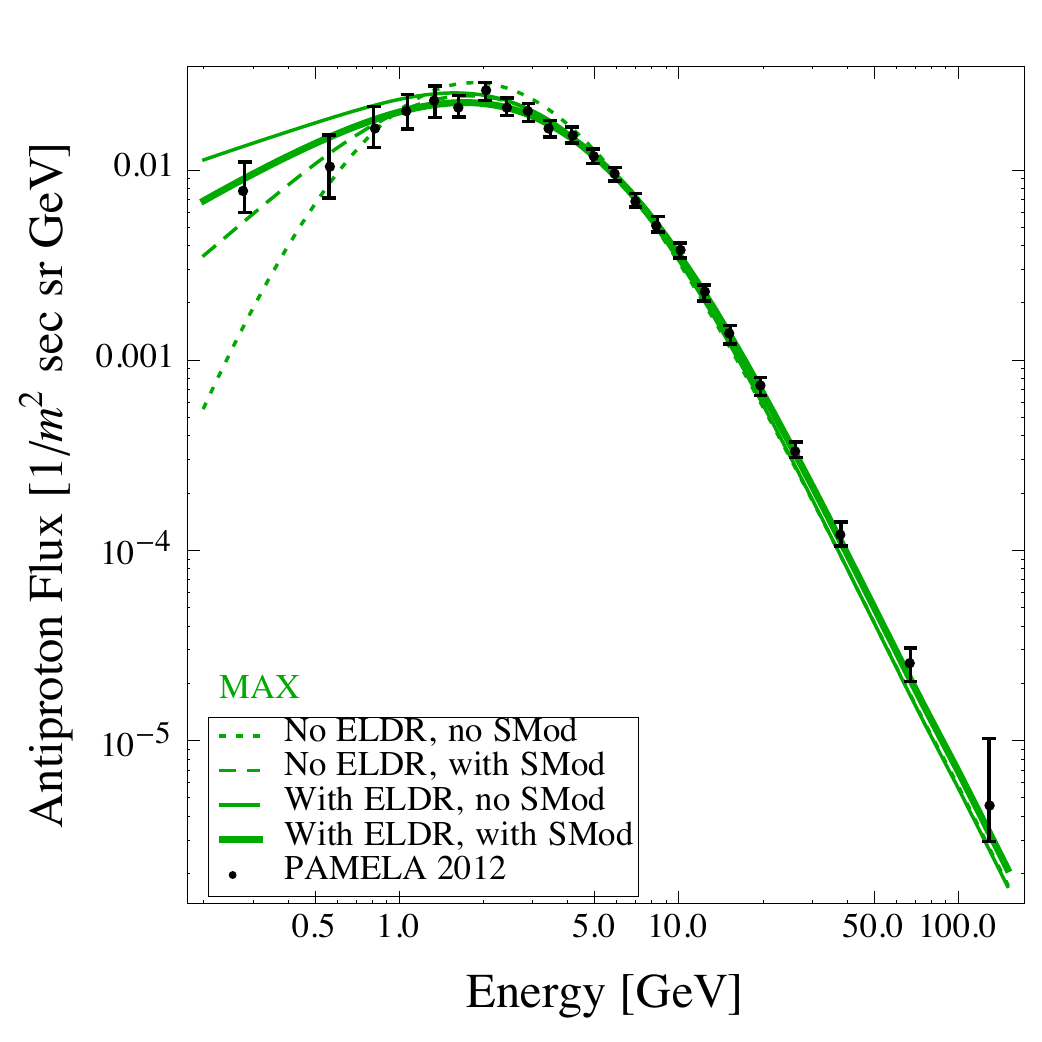}
\caption{\small \em\label{fig:bestfisk}
The {\bfseries astrophysical background (secondary) antiproton spectra} with/without SMod and with/without ELDR,
for the MIN, MED, MAX propagation models (left to right). SuperimposeD are the data points from \PAMELA.
}
\end{center}
\end{figure}

\medskip
In fig.~\ref{fig:bestfisk}, we plot the fluxes that we obtain for the MIN, MED and MAX models in order to compare now quantitatively these astrophysical 
$\bar p$ fluxes with Pamela 2012 data~\cite{Adriani:2012paa}.
%
%
To better assess the impact of our additional effects, three cases are considered:
first, we consider only the data points with kinetic energies $K>10$ GeV and we neglect solar modulation;
second, we add solar modulation (allowing the Fisk potential to vary in the range $0.1\ {\rm GV} <\phi_F<1.1$ GV and the normalization
$A$ within 1 and 1.25);
third, we use the whole energy spectrum (still taking solar modulation into account).
For each one of these cases, we compare the $\chi^2$ with and without ELDR. In table \ref{tab:bestfisk}, we present the best fit Fisk potential $\phi_F$
and the corresponding $\chi^2$ value.
In general, the agreement between the astrophysical fluxes and the data, that we achieve by including ELDR and SMod, is very good,
with a value of the reduced $\chi^2$ which is below 1 for all configurations as soon as SMod is included. This is significant and essentially implies
that little room is left for exotic contributions such as those from DM. A few fine features are however worth noticing.  

\noindent (i) Taking into account solar modulation improves the fits considerably even above 10 GeV, especially for the MIN and MED configurations.
This is because the datapoints between 10 and 50 GeV have such an exquisite precision that even the limited modification of solar modulation above 10 GeV
has an impact. 

\noindent (ii) Adding ELDR has a limited impact above 10 GeV (improving or worsening the $\chi^2$ only by fractions of a point in most cases). However,
when the whole spectrum is considered, then ELDR allows a significantly better agreement.
\begin{table}[h!]
\scriptsize
\center
\begin{tabular}{|l|c|c|c|c|c|c|c|c|c|}
\hline
& \multicolumn{3}{c|}{MIN}& \multicolumn{3}{c|}{MED}& \multicolumn{3}{c|}{MAX}\\
& $\chi^2/dof$&$A$ [-]&$\phi_F$ [GV]& $\chi^2/dof$&$A$ [-]&$\phi_F$ [GV]& $\chi^2/dof$&$A$ [-]&$\phi_F$ [GV]\\
\hline
\bf{$\bf{K>10}$ GeV no SMod}& &&&&&&&&\\
No ELDR& 35.07/7&1.00&-& 12.85/7&1.0&-&2.21/7&1.21&-\\
With ELDR&36.43/7&1.00&-&15.44/7&1.0&-&2.19/7&1.20&-\\
\hline
\bf{$\bf{K>10}$ GeV with SMod}& &&&&&&&&\\
No ELDR& 3.07/6&1.14&1.10 & 1.77/6&1.22&1.05 & 1.44/6&1.25&0.17\\
With ELDR& 3.29/6&1.25&0.79 &2.24/6&1.25&0.69 & 1.70/6&1.25&0.26\\
\hline
\bf{Whole spectrum with SMod}& &&&&&&&&\\
No ELDR& 15.63/21&1.00&0.66 & 8.23/21&1.12&0.74 & 10.16/21&1.25&0.46\\
With ELDR& 9.65/21&1.22&0.74 & 6.95/21&1.24&0.70 & 6.38/21&1.25&0.38\\
\hline
\end{tabular}
\caption{\small\em\label{tab:bestfisk}
{\bfseries $\chi^2$ of the astrophysical $\pbar$ flux} with respect to \PAMELA\ 2012 data, with and without ELDR, in three cases.
We also report the best fit values for the addition parameters $A$ and $\phi_F$, where applicable.
} 
\end{table}
%

\medskip

To conclude this analysis, we provide some useful approximating functions to the astrophysical fluxes presented above. They read:
\beq
\log \left( \frac{\Phi_{\rm bkg}}{1/{\rm m}^2 {\rm \  sec\  sr\  GeV}} \right)= \sum_{i=0}^{8} a_i  \log^i \left( \frac{K}{{\rm GeV}} \right),
\label{eq:approx_background}
\eeq
with the coefficients $a_i$ as presented in Table~\ref{tab:fittingbkg} for MIN, MED and MAX propagation models.
These functions reproduce our results within 5$\%$ in the range of energies $0.2\ {\rm GeV} \leq K \leq 800$ GeV.
They remain accurate within 20$\%$ all the way up to $10\ {\rm TeV}$.
For the convenience of the reader, we also provide an approximating function (labelled `average' in Table~\ref{tab:fittingbkg}, not to be confused with MED)
which corresponds to a flux that sits in the middle of the highest and lowest astrophysical fluxes, at all energies. We stress that of course it is preferable to use the functions adapted to the MIN, MED or MAX scenarios, as applicable, rather than this `average' function. Nevertheless, the latter can allow for a
quick scenario-independent estimate. In that case, the uncertainty due to the propagation can be taken into account as follows: for energies in the range
$0.1\ {\rm GeV} \leq K \leq 450$ GeV, the uncertainty is of 30$\%$ above and below; up to 1.7 TeV the uncertainty is 50$\%$ and it reaches $\sim 70 \%$ at
10 TeV.
%
\begin{table}[h!]
\center
\small
\begin{tabular}{|l|c|c|c|c|c|c|c|c|c|}
\hline
& $a_0$&$a_1$&$a_2$&$a_3$&$a_4$&$a_5$&$a_6$&$a_7$&$a_8$\\
\hline
MIN&-1.5116 & 0.37991 & - 0.69686 &- 0.92051 &  0.17873 & 0.34161 & - 0.19910 & 0.04154 & - 0.00309\\
MED&-1.5079 & 0.34279 & - 0.71347 & - 0.89053 & 0.18914 & 0.32518 & - 0.19223 & 0.04022 & - 0.00299 \\
MAX&-1.6192 & 0.34986 &- 0.69943 & - 0.86758 & 0.19016 &  0.30858 & - 0.18192 &  0.03777 & - 0.00279 \\
\hline
\\
\hline
`average' &-1.5594 & 0.35685 & - 0.69211 & - 0.89153 & 0.17990 & 0.32460 & - 0.18875 & 0.03911& - 0.00289 \\
\hline
\end{tabular}
\caption{\small\em\label{tab:fittingbkg}
{\bfseries Coefficients for the approximating functions} of the astrophysical flux, assuming $A=1$ and
$\phi_F=0$ for MIN, MED and MAX propagation scenarios and for the `average' case.}
\end{table}

%

\bigskip

Recently, refs.~\cite{Kappl:2014hha} and~\cite{Hooper:2014ysa} have revisited the computation of the secondary antiproton flux. (The latter has also obtained new antiproton constraints on Dark Matter annihilation, on which we will comment later.) Our results essentially agree with the findings of the former so we mostly comment here on the latter. Such work bases its computation on a large set of propagation parameter
determined in~\cite{Trotta:2010mx}, via a full bayesian scan of cosmic ray data, and its crucial result is that the secondary $\bar p$ flux is found to
systematically undershoot the $\bar p$ data by \PAMELA\ (the resulting `antiproton excess' can then be fit with a DM contribution as discussed in
ref.~\cite{Hooper:2014ysa}). Similar results had been anticipated in previous works~\cite{otherpbar}. However, other set of studies~\cite{otherpbar2,Kappl:2014hha}
do not reach the same conclusion and, in particular, this is at odd with what we find: as discussed, we obtain a secondary flux which is well in agreement
with \PAMELA\ data for any one of the propagation models we consider and essentially across the whole range of energies (see e.g. fig.~\ref{fig:bestfisk}).
There might be many origins for this difference and it is difficult to pin down any single one, as discussed in previous works and as the authors
of~\cite{Hooper:2014ysa} also very nicely point out. Firstly, we notice that ref.~\cite{Hooper:2014ysa} considers two values of the Fisk potential for
antiprotons which are smaller or equal to the value for protons, while we do not restrict to this situation. Actually, as already mentioned above,
computations based on~\cite{Maccione:2012cu} show a preference for $\phi_F^{\bar p} > \phi_F^{p}$ during the \PAMELA\ data taking
period~\cite{Gaggero}. While in first approximation a large $\phi_F^{\bar p}$ would tend to reduce the $\bar p$ spectrum further, it is possible
that the interplay with other effects mentioned below produces the opposite effect. Secondly, the models considered in~\cite{Hooper:2014ysa},
with one exception, do not include convection. Adding convection would generically have the effect of reducing an antiproton excess, as also
recognized in~\cite{Hooper:2014ysa}~\footnote{Moreover, position-dependent convection and diffusion can have a very interesting impact~\cite{Gaggero:2014xla}.}. Thirdly, the primary proton spectrum considered in~\cite{Hooper:2014ysa} (a broken power law with a break
at 10 GV) differs from the one we use based on \PAMELA\ data. Finally, we introduce an additional degree of freedom
(linked to the antineutron production, as discussed above) which affects the overall normalization of the $\bar p$ flux and which is not taken into 
account in~\cite{Hooper:2014ysa}. It is conceivable that a combination of these individually small effects works in the direction of explaining
away the difference.

In any case, the comparison with the detailed work in~\cite{Hooper:2014ysa} shows how crucial is the impact of the `fine points' in the calculation
of the secondaries. We will have to keep this in mind and proceed with maximal caution.

\subsection{The dark matter signal}
\label{sec:antip_DM}

We here briefly review the antiproton fluxes from Dark Matter, referring the reader to~\cite{PPPC4DMID} for more details. 

Primary antiprotons originate from DM annihilations or decays in each point of the galactic halo. Hence they constitute, for the purposes of the transport equation (\ref{eq:transport}), a source term $Q$ which reads 

\beq 
Q^{\rm prim}_{\bar p} = \frac{1}{2} \left(\frac{\rho}{M_{\rm DM}}\right)^2 f^{\rm ann}_{\rm inj},\qquad f^{\rm ann}_{\rm inj} = \sum_{f} \langle \sigma v\rangle_f \frac{dN_{\bar p}^f}{dK} \qquad {\rm (annihilation)},
\label{eq:Qann}
\eeq
\beq 
Q^{\rm prim}_{\bar p} = \left(\frac{\rho}{M_{\rm DM}}\right) f^{\rm dec}_{\rm inj},\qquad f^{\rm dec}_{\rm inj} = \sum_{f} \Gamma_f \frac{dN_{\bar p}^f}{dK} \qquad {\rm (decay)}.
\label{eq:Qdec}
\eeq
The above formul\ae\ show the well known factorization of the source term in a portion that depends essentially on astrophysics (the DM density distribution $\rho$, for which we discuss typical choices below) and in a portion ($f_{\rm inj}^{\rm ann/dec}$) that depends on the particle physics model. Here $dN_{\bar p}/dK$ are the antiproton spectra per single annihilation or decay event and $f$ runs over all the channels with $\bar p$ in the final state, with the respective thermal averaged cross sections $\sigma v$ or decay rate $\Gamma$.

\subsubsection{Particle physics input}
\label{sec:ppinput}

Following~\cite{PPPC4DMID}, we consider a complete array of annihilation or decay channels, in a model independent way. They consist in the following 2 $\times$ 23 cases:
\beq 
\left.
\begin{array}{rr}
{\rm annihilation} & {\rm DM} \ {\rm DM}  \\
{\rm decay} & {\rm DM}
 \end{array}\right\} \to
\left\{
\begin{array}{ll}
e_L^+e_L^-,\ 
e_R^+e_R^-,\
\mu_L^+\mu_L^-,\ 
\mu_R^+\mu_R^-,\ 
\tau_L^+\tau_L^-,\ 
\tau_R^+\tau_R^-,\\[3mm]
q \bar q, \ 
c \bar c, \ 
b \bar b, \  
t \bar t, \ 
\gamma \gamma,\ 
g g,  \\[3mm]
W_L^+ W_L^-,\ 
W_T^+W_T^-,\ 
Z_LZ_L,\ 
Z_T Z_T, \\[3mm]
hh, \\[3mm]
\nu_e \bar\nu_e, \ 
\nu_\mu \bar\nu_\mu, \ 
\nu_\tau \bar\nu_\tau, \\[3mm] 
VV \to 4e, \
VV \to 4\mu, \
VV \to 4\tau.
\end{array}
\right.
\label{primarychannels}
\eeq
Here the subscript $_{L,R}$ denote the $L$eft-handed and $R$ight-handed polarizations for the leptons and $_{L,T}$ the $L$ongitudinal or $T$ransverse ones for the gauge boson: since ElectroWeak corrections (mentioned in the Introduction) act differently on the different polarizations, it is important to keep them separate. $q ={u,d,s}$ denotes a light quark and $h$ is the Standard Model (SM) Higgs boson, with its mass fixed at 125 GeV.
The last three channels denote models in which the annihilation or decay first happens into some new (light) boson $V$ which then decays into a pair of leptons. 

As for the DM mass, we consider the range $m_{\rm DM} = 5\, {\rm GeV} \to 100\, {\rm TeV}$ (annihilation) or $m_{\rm DM} = 10\, {\rm GeV} \to 200\, {\rm TeV}$ (decay).
For additional details, we refer to~\cite{PPPC4DMID}. 

\subsubsection{Astrophysical parameters}

%
\begin{table}[t]
\begin{center}
\begin{tabular}{l|crc}
 &  \multicolumn{3}{c}{DM halo parameters}  \\
DM halo & $\alpha$ &  $r_{s}$ [kpc] & $\rho_{s}$ [GeV/cm$^{3}$]\\
  \hline 
  NFW~\cite{Navarro:1995iw} & $-$ & 24.42 & 0.184 \\
  Einasto~\cite{Graham:2005xx, Navarro:2008kc} & 0.17 & 28.44 & 0.033 \\
  EinastoB & 0.11 & 35.24 & 0.021 \\
  Isothermal~\cite{Begeman,Bahcall:1980fb} & $-$ & 4.38 & 1.387 \\
  Burkert~\cite{Burkert} & $-$ & 12.67 & 0.712 \\
  Moore~\cite{Moore04} & $-$ & 30.28 & 0.105 
 \end{tabular}
 \end{center} 
\caption{\em \small The {\bfseries Dark Matter profile parameters} to be plugged in the functional forms of eq.~(\ref{eq:profiles}). These specific values are derived as discussed in~\cite{PPPC4DMID}. 
\label{tab:profileparam}}
\end{table}
%

Following again~\cite{PPPC4DMID}, we consider the following various DM halo profiles~\footnote{See also~\cite{Nesti:2013uwa} for a more recent discussion.}:
\begin{equation}
\begin{array}{rrcl}
{\rm NFW:} & \rho_{\rm NFW}({\mathbf r})  & = & \displaystyle \rho_{s}\frac{r_{s}}{{\mathbf r}}\left(1+\frac{{\mathbf r}}{r_{s}}\right)^{-2} \\[4mm]
{\rm Einasto:} & \rho_{\rm Ein}({\mathbf r})  & = & \displaystyle \rho_{s}\exp\left\{-\frac{2}{\alpha}\left[\left(\frac{{\mathbf r}}{r_{s}}\right)^{\alpha}-1\right]\right\} \\[4mm]
{\rm Isothermal:} &  \rho_{\rm Iso}({\mathbf r}) & = & \displaystyle \frac{\rho_{s}}{1+\left({\mathbf r}/r_{s}\right)^{2}} \\[4mm]
{\rm Burkert:} & \rho_{\rm Bur}({\mathbf r}) & = & \displaystyle  \frac{\rho_{s}}{(1+{\mathbf r}/r_{s})(1+ ({\mathbf r}/r_{s})^{2})} \\[4mm]
{\rm Moore:} &  \rho_{\rm Moo}({\mathbf r})  & = & \displaystyle \rho_{s} \left(\frac{r_s}{{\mathbf r}}\right)^{1.16} \left(1+\frac{{\mathbf r}}{r_s}\right)^{-1.84} .
\end{array}
\label{eq:profiles}
\end{equation} 
Here ${\mathbf r}$ is the galactocentric distance and spherical symmetry is always assumed. 
The values for the parameters $\rho_s$, $r_s$ and $\alpha$ are presented in table~\ref{tab:profileparam}.
The profiles span a very wide range of possibilities, from the very peaked profiles such as `Moore' to the cored profiles such as `Burkert' or `Isothermal'. The `EinastoB' profile, introduced in~\cite{PPPC4DMID}, features a smaller $\alpha$ value which translates into a larger density towards the Galactic Center (with respect to `Einasto'), describing the  possible contraction due to the inclusion of baryons in the simulations. 

With the values in table~\ref{tab:profileparam}, all profiles are normalized to give a local DM density of 0.3 GeV$/{\rm cm}^3$.
The distance of the Sun to the Galactic Center is taken to be 8.33 kpc on the basis of~\cite{rSun} and consistently with~\cite{PPPC4DMID}.

\subsubsection{DM antiproton spectra including `ELDR'}
\label{sec:DMspectra}

\begin{figure}[t!]
\begin{center}
\includegraphics[width=0.32\textwidth]{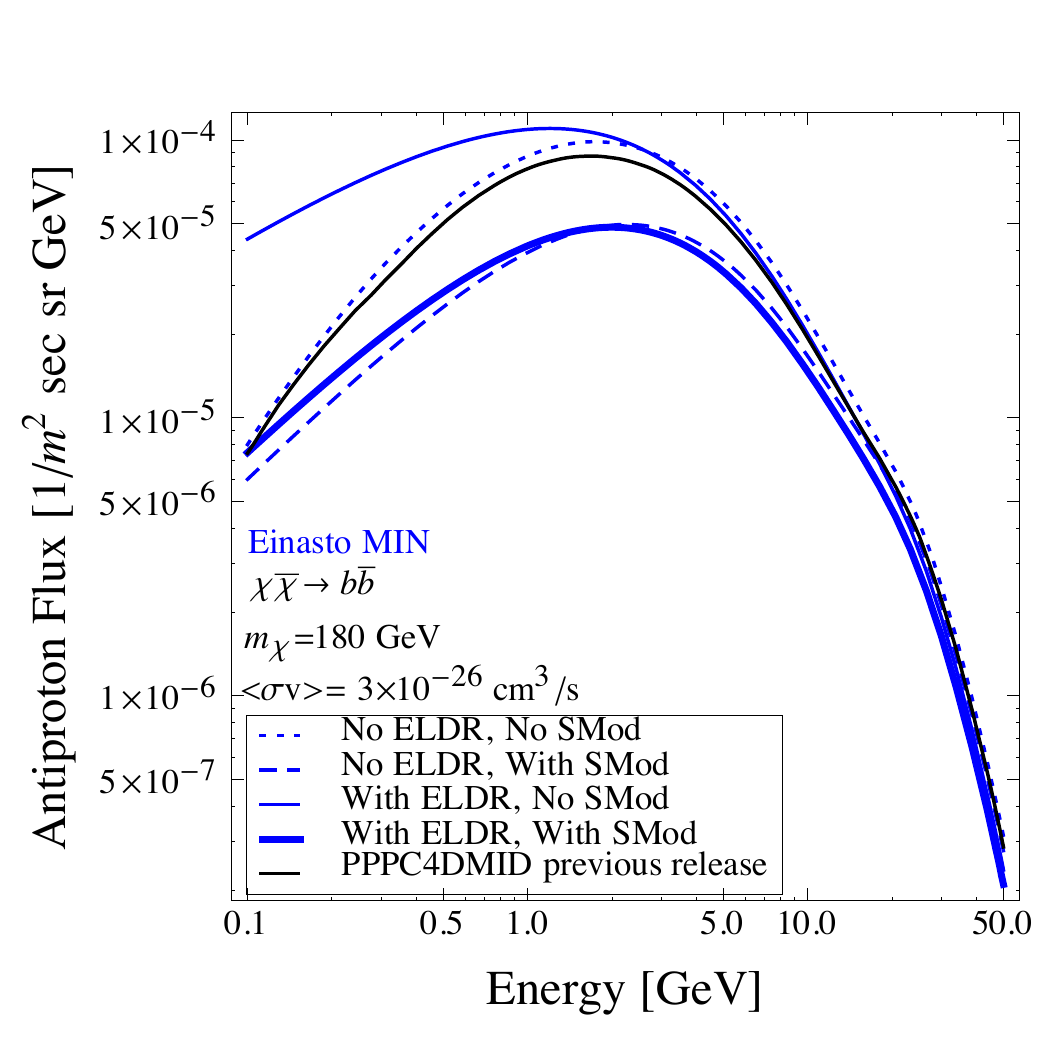}
\includegraphics[width=0.32\textwidth]{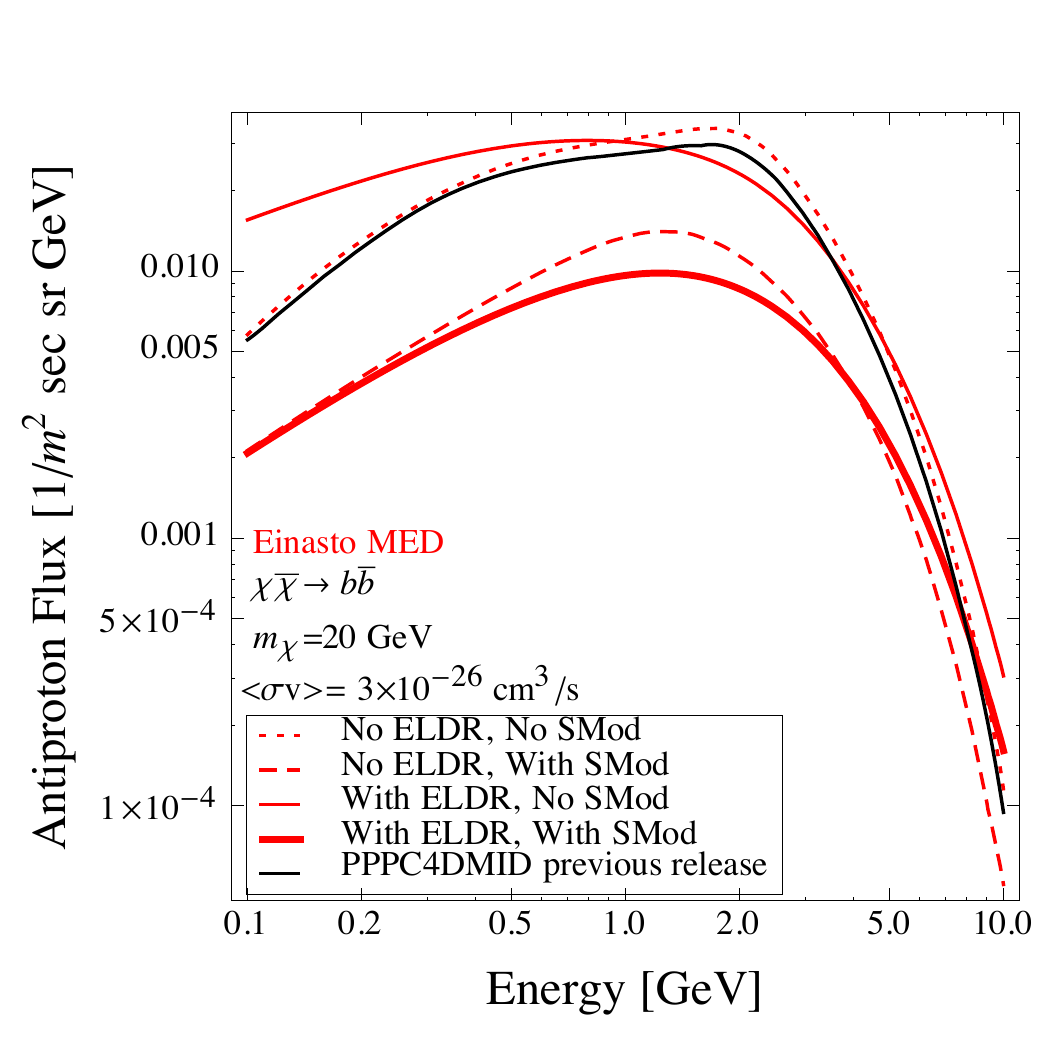}
\includegraphics[width=0.32\textwidth]{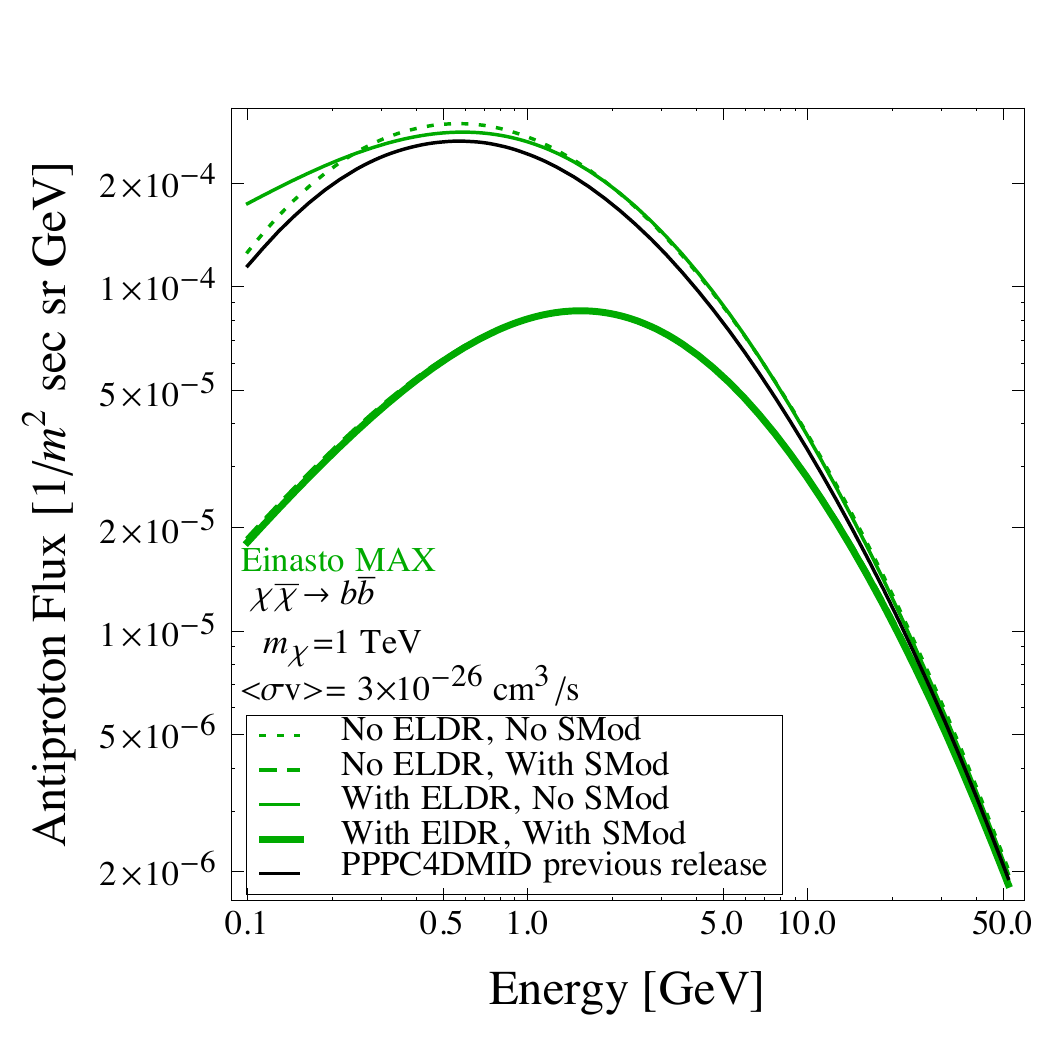}
\caption{\small \em\label{fig:primary_spectra} Comparison of the {\bfseries antiproton flux from Dark Matter} annihilating into $b\bar{b}$ for an Einasto profile with/without SMod, with/without ELDR and with the previous {\sc Pppc4dmid} flux, for the three propagation models MIN, MED, MAX and for three chosen masses.}
\end{center}
\end{figure}

Fig.~\ref{fig:primary_spectra} presents the DM antiproton spectra (for some representative choices of masses) computed and propagated as discussed in the previous sections.
We show the spectra from the previous release of {\sc Pppc4dmid} (Release 3.0) compared with the spectra from our current calculation, with and without ELDR. 
The previous {\sc Pppc4dmid} and the new calculation without ELDR agree very well (with small residual differences due to the different computational techniques), which is expected and reassuring.
Including ELDR, the main effect is the one of having the spectrum `squat': the peak decreases, the low energy tail is softened and the high energy portion can be somewhat raised. It is interesting to note that this leads also to a non-zero $\bar p$ flux {\em above the nominal endpoint} of the spectrum at $K = m_{\rm DM}$, thanks to the (re)acceleration experienced by some of those antiprotons which had been produced with energies already close to such endpoint.

\bigskip

The antiproton spectra including ELDR are our most refined output possible for this observable and constitute our final result. We compute them for all the channels and range of masses spelled out in sec.~\ref{sec:ppinput} and we put them at public disposal in the new release of \href{http://www.marcocirelli.net/PPPC4DMID.html}{{\sc Pppc4dmid}} (Release 4.0).
Solar modulation, being inherently epoch-dependent (see sec.~\ref{sec:SolarMod}), is not included in the numerical product, but can be easily implemented with the use of eq.~(\ref{eq:SMod}).
%
%

\section{Constraints on Dark Matter}
\label{sec:constraints}

With the astrophysical background discussed in section \ref{sec:antip_background} and the fluxes from dark matter annihilation obtained in section \ref{sec:antip_DM}, we can now compute the constraints on Dark Matter in the usual `mass vs. annihilation cross section' plane. We do not aim here at an exhaustive scan of annihilation channels and DM profiles, but rather to show the impact on some specific examples. For definiteness, we focus on annihilations into $\bar b b$ and an Einasto profile. 

\medskip

At a practical level, for each propagation model, we add the contributions of the astrophysical flux and the DM, so that the total flux is
\begin{equation}
\Phi_{\rm tot} (m_{\rm DM}, \langle \sigma v \rangle, \phi_F)=\Phi_{\rm bkg}(\phi_F)+\Phi_{\rm DM}(m_{\rm DM}, \langle \sigma v \rangle, \phi_F),
\end{equation}
where $\phi_F$ is the Fisk potential over which we marginalize. Then, for each mass $m_{\rm DM}$, we solve the following equation in $\langle \sigma v \rangle$
\begin{equation}
\chi^2_{\rm DM}(m_{\rm DM},\langle \sigma v \rangle, \phi_F)-\chi^2_0=4,
\end{equation}
where $\chi^2_0$ is the chi square of the best fit background, found in table \ref{tab:bestfisk}.
\\

\subsection{Current constraints using \PAMELA\ 2012 data}
\label{sec:current_constraints}

\begin{figure}[t!]
\begin{center}
\includegraphics[width=0.328\textwidth]{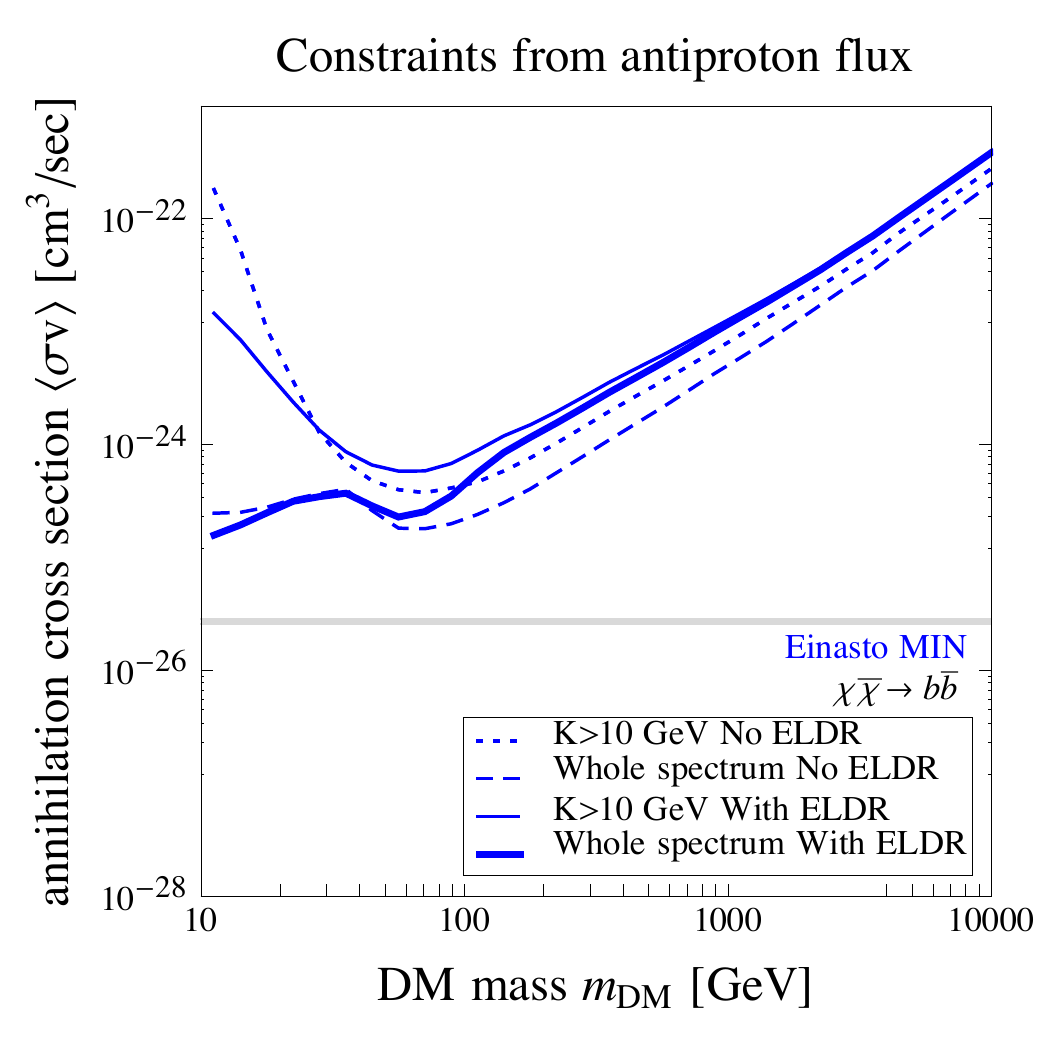}
\includegraphics[width=0.328\textwidth]{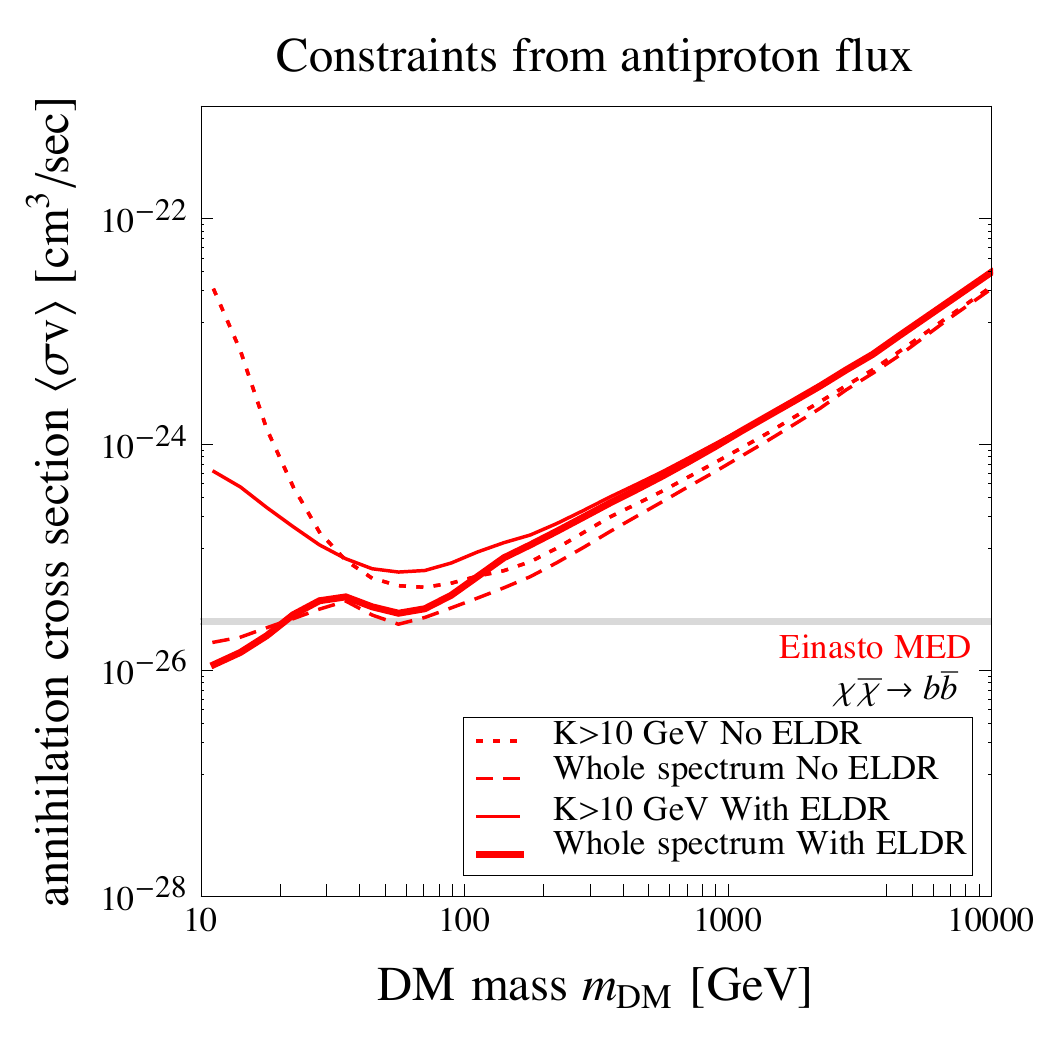}
\includegraphics[width=0.328\textwidth]{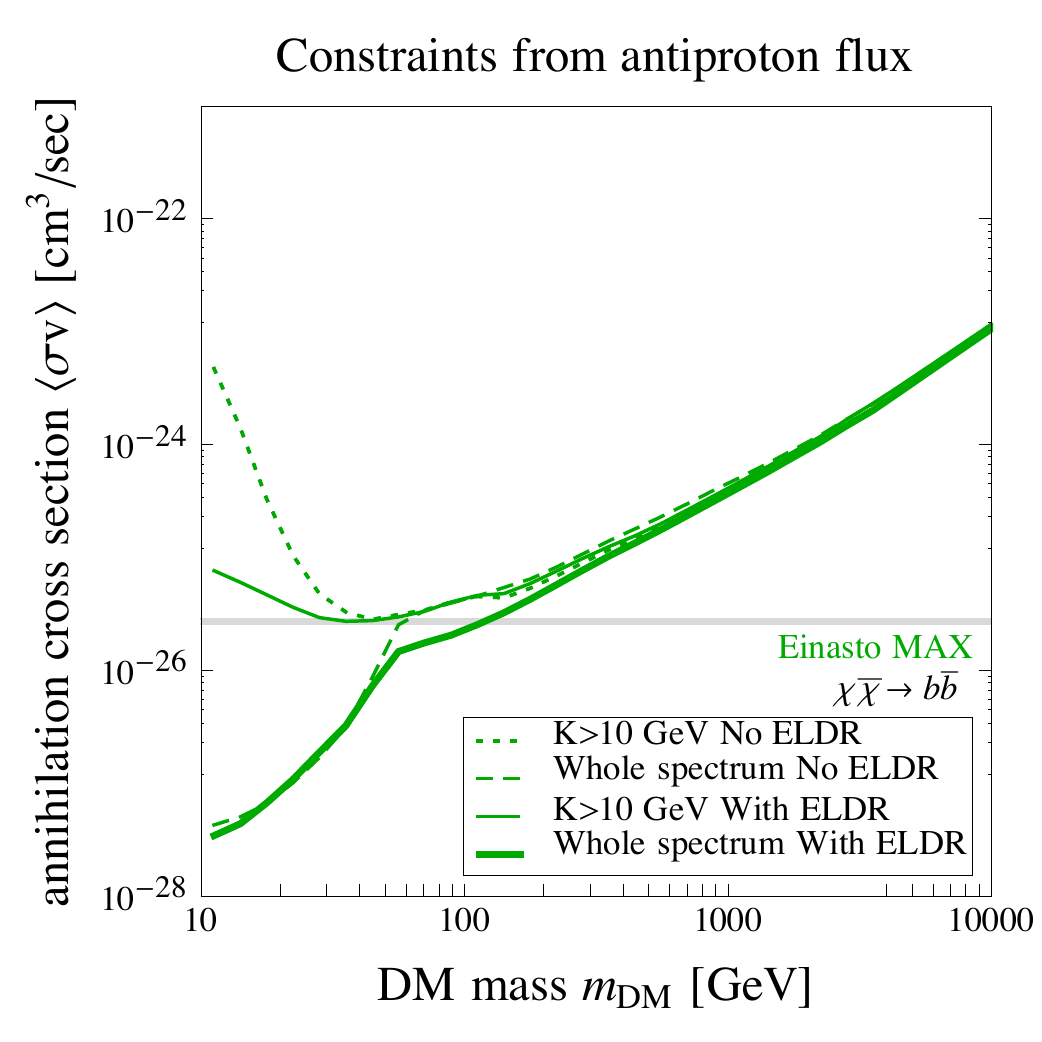}
\caption{\small \em\label{PAMELAcontraints} {\bfseries Antiproton constraints} on Dark Matter annihilating into $b\bar{b}$ for an Einasto profile with or without ELDR, for $K>10 \GeV$ or the whole spectrum and for the three propagation models MIN, MED, MAX. Solar modulation is marginalized over as explained in the main text.}
\end{center}
\end{figure}

\begin{figure}[t!]
\begin{center}
\includegraphics[width=0.5\textwidth]{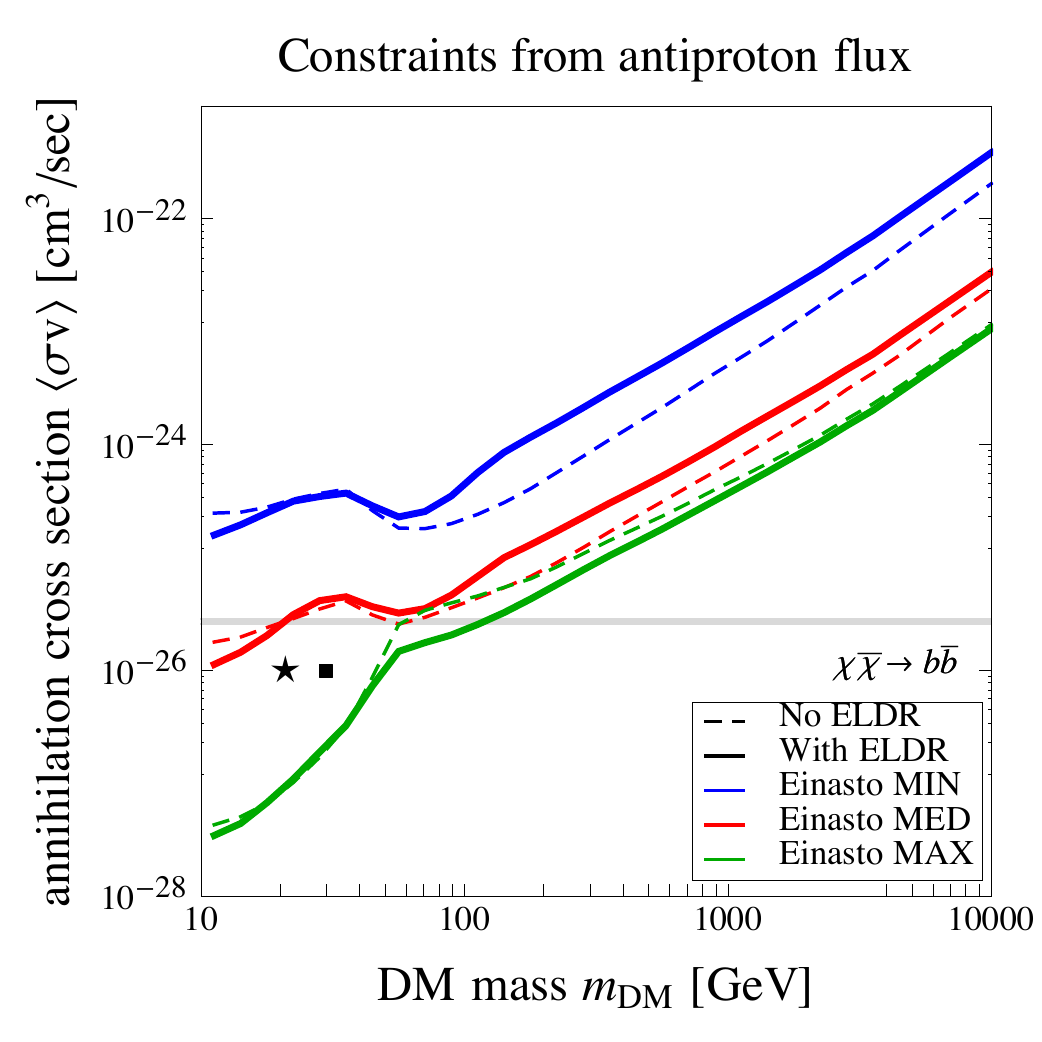}
\caption{\small \em\label{PAMELAconstraints2} {\bfseries Summary of the antiproton constraints} on Dark Matter annihilating into $b\bar{b}$ for an Einasto profile with or without ELDR for the three propagation models MIN, MED, MAX.
The symbols represent the DM cases listed in Table \ref{tab:Deltachi2}.}
\end{center}
\end{figure}

The results obtained with data from \PAMELA\ 2012~\cite{Adriani:2012paa} are presented in figure~\ref{PAMELAcontraints}. For each propagation model, we distinguish four different cases: taking into account only the data points with energies $K>10$ GeV or for the whole spectrum and with or without ELDR. In fig.~\ref{PAMELAconstraints2} we keep only the `whole spectrum' case and compare the three propagation scenarios.

For the MIN and MED propagation models, we observe that the constraints with ELDR are stronger at small masses and weaker at large masses than the ones obtained without these effects. In fact, as we have seen in section \ref{sec:ELDR}, including ELDR means depopulating states with high energies and adding them to lower energies.
Thus, with ELDR the astrophysical background as well as the DM signal are lower at high energies leaving more freedom for a large mass DM contribution and relaxing the constraints.
At low masses, the situation is the opposite: the astrophysical background and the DM signal are higher with ELDR and thus the constraints are stronger.
For MAX, the effect is almost absent or actually inverted. This can be qualitatively understood in the light of the discussion of the relative importance of reacceleration and energy losses presented in Sec.~\ref{sec:ELDR}. 
In short, while for MIN and MED the effect of energy losses is dominant at high energies, for MAX it is counterbalanced by diffusive reacceleration, the astrophysical and DM spectra slightly increase and hence the constraints are faintly stronger.

The other prominent feature in the bounds is the `hump' below $m_{\rm DM} \sim$40 GeV, especially visible for the MIN and MED cases. It originates from another rather complex interplay, this time among the effect of SMod, the size of the experimental error bars and the shape of the spectra. Indeed, for DM masses of the order of 20$-$40 GeV, the DM $\bar p$ spectrum peaks at the same energy ($\approx 2$ GeV) and has the same shape as the astrophysical background. By playing with an appropriate choice of the Fisk potential, more room can be freed for the DM, hence the constraints relax. Above $\sim 40$ GeV, instead, the DM component mostly contributes to data points above $\sim$ 5 GeV: here the error bars are smaller and the SMod effect cannot effectively act as a compensation to DM, hence the constraints are more stringent.

\medskip

%

Before moving on, we comment on another subtle result, which is not directly visible in the plots of fig.~\ref{PAMELAcontraints} but which can be inferred by closely scrutinizing the $\chi^2$ of the fits: the inclusion of ELDR is important in the computation of $\bar p$ fluxes from low mass DM because, without it, a DM signal can even slightly improve the $\chi^2$ and therefore be artificially favored. We show two specific, cherry-picked examples in table \ref{tab:Deltachi2}. Focussing for definiteness on the first case (the 20 GeV DM), we see that: when ELDR is correctly included, the fit with respect to the one of a pure astrophysical background is very slightly worsened ($\Delta \chi^2 = 0.04$ in this example); if instead ELDR effects are neglected, the fit can be ameliorated (by 3.78 points in this example), inducing an unfounded preference in favor of DM. 
Qualitatively, the origin of this effect can be understood by looking at the best fit background in figure \ref{fig:bestfisk}: without ELDR, the astrophysical component typically passes just below the first few data points; adding a DM contribution can push the curve up and improve the $\chi^2$. This does not happen when ELDR is added, as the astrophysical background fits the data better at these low energies. We will come back to this issue in more detail when discussing \AMS\ projections in the next section.

 \begin{table}[h!]
\small
\center
\begin{tabular}{|c|c|c|c|c|c|c|c|c|c|}
\hline
\rule{0pt}{2.6ex} Symbol&$m_{\rm DM}$ [GeV]& $\langle \sigma v \rangle$ [cm$^3$s$^{-1}$] & \multicolumn{3}{c|}{No ELDR}& \multicolumn{3}{c|}{With ELDR}  \\
\rule[-1.2ex]{0pt}{0pt} &&&A[-] &$\phi_F$ [GV]&$\Delta \chi^2$&A[-]&$\phi_F$ [GV]&$\Delta \chi^2$\\
\hline
$\bigstar$&20&$10^{-26}$&1.20&0.98&-3.78&1.25&0.86&0.04 \rule{0pt}{2.6ex} \\
\scriptsize$\blacksquare$&30&$10^{-26}$&1.15&0.90&-1.66&1.25&0.82&0.17\\

\hline
\end{tabular}
\caption{\small\em\label{tab:Deltachi2} Examples of the $\Delta \chi^2$ of the fit to \PAMELA\ data with respect to a pure background case obtained by adding a DM contribution (with the specified parameters), with or without ELDR in the MED propagation model.} 
\end{table}
%


\medskip

Antiproton constraints based on \PAMELA\ data like the ones that we obtain here have also been deduced recently in~\cite{Cirelli:2013hv},~\cite{Bringmann:2014lpa},~\cite{hooperon} and \cite{Hooper:2014ysa}. Our constraints are consistent with~\cite{Cirelli:2013hv} (except that that work had not included ELDR) and essentially with~\cite{hooperon} too, while the constraints in~\cite{Bringmann:2014lpa} are more stringent than ours and those in \cite{Hooper:2014ysa} are much looser. While a detailed case-by-case comparison is difficult, the choice of the astrophysical background (on which we commented extensively in sec.~\ref{sec:antip_background} in connection with~\cite{Hooper:2014ysa}) plays probably the major role in explaining the differences.

\medskip

As a rather general conclusion, the different aspects of our analysis and the comparisons with other results show how critical the fine details of background and signal fluxes (ELDR, SMod, interplay with the propagation setups) are to establish the constraints. These should be therefore handled with great care.

\subsection{Future Sensitivity of \AMS}
\label{sec:future_sensitivity}

In this section we assess what  will be the future sensitivity to DM annihilation of the antiproton measurements by the \AMS\ experiment. 

\begin{figure}[t!]
\begin{center}
\includegraphics[width=0.475\textwidth]{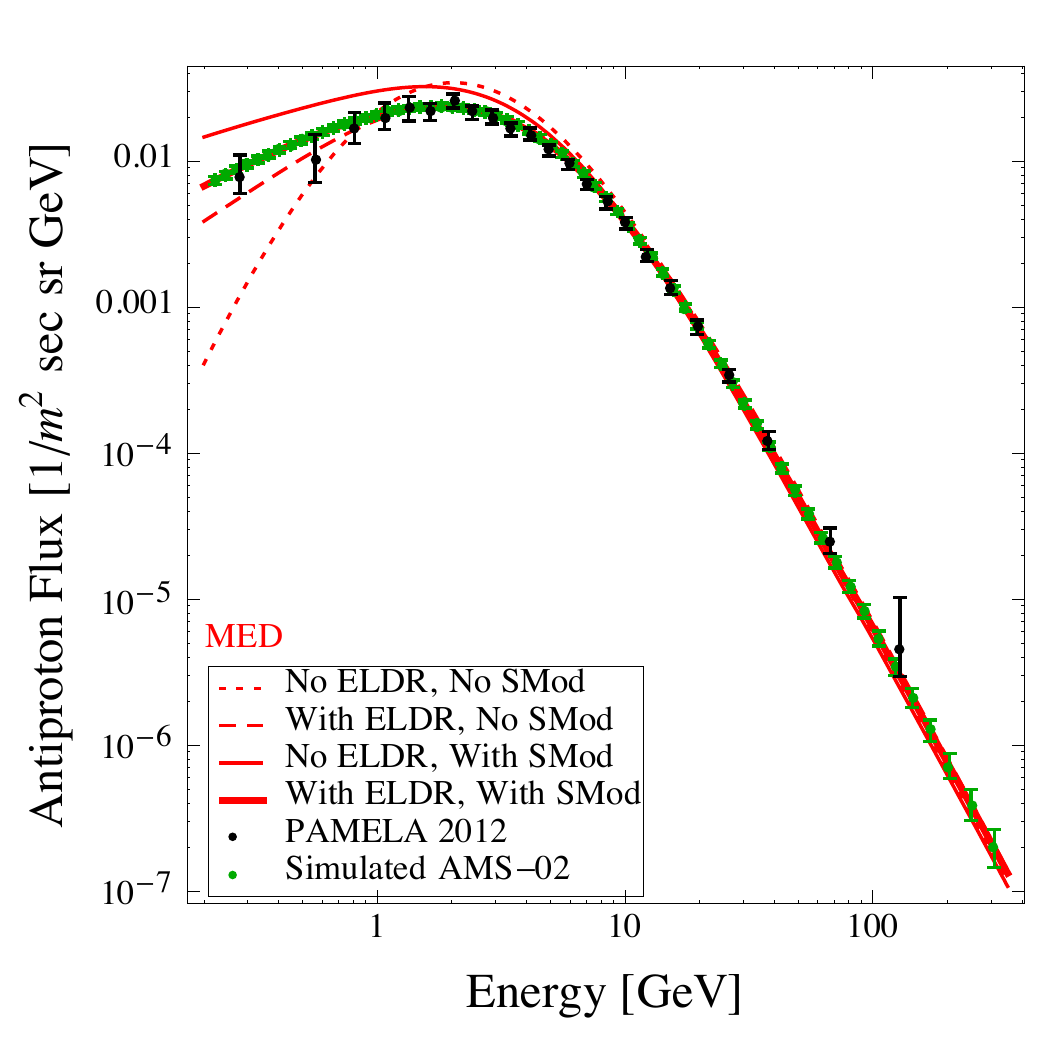} \quad
\includegraphics[width=0.49\textwidth]{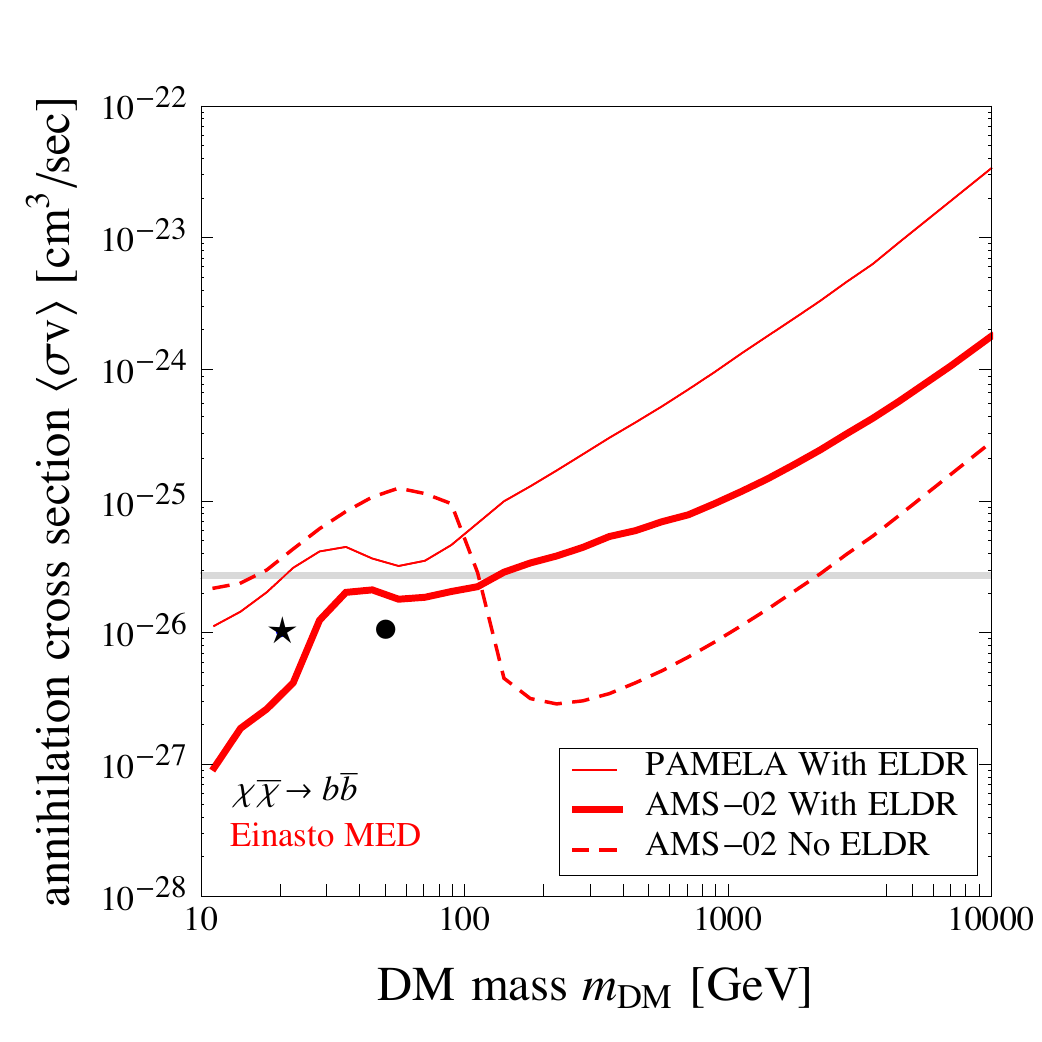}
\caption{\small \em\label{AMSconstraints} {\bfseries Predictions of \AMS\ sensitivity}. Left: mock data points for the $\bar p$ flux measured by \AMS\ after 1 year of data-taking and for a Fisk potential $\phi_F=0.6$ GV. Right: Sensitivity of \AMS\ to Dark Matter annihilating into $b\bar{b}$ for an Einasto profile with or without ELDR, for the MED propagation model. The symbols represent the DM cases listed in Table \ref{tab:Deltachi2AMS}.}
\end{center}
\end{figure}

We produce simulated \AMS\ data by putting the points on the MED curve of the astrophysical background, computed including ELDR and a Fisk potential $\phi_F=0.6$. The binning in energy and the error bars are computed in the same way as in~\cite{Cirelli:2013hv}. The mock data points obtained in this way are presented in figure~\ref{AMSconstraints}. We then apply the same fitting procedure that we used in the previous section, including in particular the marginalization over the value of the Fisk potential.

\medskip

First, we check the quality of the fits with background only. The results are shown in table \ref{tab:chi2_AMS}. We correctly find that $\chi^2=0$ for the MED case with ELDR and the correct Fisk potential is recovered. The fit remains very good also for MIN and MAX, at the condition of adjusting the Fisk potential to a respectively larger/smaller value than the true one. If ELDR are neglected, the $\chi^2$ worsens dramatically. This is not surprising given the small error bars of the mock data: the best fit curves undershoot the data for the first 15 points and overshoots them at high energies.

 \begin{table}[h!]
\small
\center
\begin{tabular}{|l|c|c|c|c|c|c|}
\hline
& \multicolumn{2}{c|}{MIN}& \multicolumn{2}{c|}{MED}& \multicolumn{2}{c|}{MAX}\\
& $\chi^2/{\rm d.o.f.}$&$\phi_F$ [GV]& $\chi^2/{\rm d.o.f.}$&$\phi_F$ [GV]& $\chi^2/{\rm d.o.f.}$&$\phi_F$ [GV]\\
\hline
\bf{Whole spectrum with SMod}& &&&&&\\
No ELDR &645/65 & 0.81&385/65 &0.74 &385/65&0.45\\
With ELDR & 19.6/65&0.62 &0/65 &0.60&30.8/65 &0.35\\
\hline
\end{tabular}
\caption{\small\em\label{tab:chi2_AMS} Fits of mock \AMS\ data with astrophysical antiprotons only, with or without ELDR.} 
\end{table}

Next, we add a DM component and compute the sensitivity of \AMS\ to DM annihilation, in the usual plane $(m_{\rm DM},\sigma v)$, on the basis of the mock data. The results are shown in figure \ref{AMSconstraints}. The behavior of the sensitivity curve is very similar to the actual limits obtained with \PAMELA\ data in the previous section. The limits at large DM masses are more constraining when ELDR are not taken into account; at low masses, it is the opposite. The curves show the `hump' at $\sim 40$ GeV already discussed in the previous section.

\medskip

Another relevant feature on which we want to comment is the significant worsening of the mock constraints for $m_{\rm DM} \lesssim 100$ GeV when ELDR are not taken into account. This arises, in this exercise we are performing, from the same mechanism we discussed in the previous section, now enhanced by the foreseen accuracy of the \AMS\ data. 
Indeed, without ELDR a pure background can not acceptably fit the data (as shown in table \ref{tab:chi2_AMS}). The $\chi^2$ can be significantly improved by introducing a DM component, hence the constraints relax. This may even lead to believe that a DM signal is hidden at low masses while instead it is just a poor modelization of the background which is at play. If fact, table \ref{tab:Deltachi2AMS} shows that a DM with a mass $m_{\rm DM}=20$ GeV (and the other parameters as listed) ameliorates the $\chi^2$ by a large amount. It sits indeed in the region allowed by the `No ELDR' curve in fig.~\ref{AMSconstraints}. By correctly including ELDR, however, the $\Delta \chi^2$ becomes positive again and the point sits in the excluded region. While the precise values of the $\chi^2$ have relatively little meaning here, being based on mock data, the point that they illustrate is valid. The general lesson, not surprisingly, is that the robustness of any DM identification in future data, especially at low energies, will have to be based on a careful understanding of the appropriate background, including in particular fine effects such as ELDR.

%

\begin{table}[h!]
\small
\center
\begin{tabular}{|c|c|c|c|c|c|c|c|c|c|}
\hline
Symbol&$m_{\rm DM}$ [GeV]& $\langle \sigma v \rangle$ [cm$^3$s$^{-1}$] & \multicolumn{3}{c|}{No ELDR}& \multicolumn{3}{c|}{With ELDR}\\
&&&$A$ [-]&$\phi_F$ [GV]&$\Delta \chi^2$&$A$ [-]&$\phi_F$ [GV]&$\Delta \chi^2$\\
\hline
$\bigstar$&20&$10^{-26}$&1.19&0.86&-342&1.25&0.78&21.6\\
\large$\bullet$&50&$10^{-26}$&1.19&0.74&-100&1.23&0.66&1.17\\
\hline
\end{tabular}
\caption{\small\em\label{tab:Deltachi2AMS} 
(Analogously to table~\ref{tab:Deltachi2}), examples of the $\Delta \chi^2$ of the fit to mock \AMS\ data with respect to a pure background case obtained by adding a DM contribution (with the specified parameters), with or without ELDR in the MED propagation model.} 
\end{table}
%

\medskip

Lastly, we note that with \AMS\ it will be possible to exclude a thermal cross section for  $m_{\rm DM}<150$ GeV. This holds using the mock data generated assuming a normalization $A=1.24$. If the real data have a smaller normalization, the constraints will be stronger and it may be possible to probe the thermal cross-section even for $m_{\rm DM}\lesssim 300$ GeV. Gamma ray searches, e.g. from dwarf galaxies, are currently probing similar ranges~\cite{dwarfs_Fermisymposium2014}, which shows how competitive the antiproton constraints can be.

\section{Conclusions}
\label{sec:conclusions}

Cosmic ray antiprotons have been regarded since long as a powerful probe for the identification of a signature of Dark Matter annihilations (or decay) in the Galaxy. Rightly so, because the physics that controls their production and propagation in the Galaxy as well as the astrophysical background are under relatively better control than for other species, and because experimental data in the antiproton channel are impressively accurate. Still, precisely for the sake of comparing with ever more accurate data, the accuracy of the theoretical predictions must be improved.
\medskip

In this work we have reassessed the computation of the astrophysical and DM antiproton fluxes by including effects such as $\bar p$ Energy Losses (including tertiary component) and Diffusive Reacceleration (`ELDR'), as well as Solar Modulation (`SMod'). These effects are often perceived as subdominant, but they can actually have an important impact, especially at low energies (hence in particular for small DM masses, $\lesssim 50$ GeV). In sec.~\ref{sec:antip_background} we have obtained the updated astrophysical background fluxes, which we provide in terms of approximating functions (eq.~(\ref{eq:approx_background}). In sec.~\ref{sec:DMspectra} we have computed the DM $\bar p$ fluxes, which we provide in the new release of \href{http://www.marcocirelli.net/PPPC4DMID.html}{{\sc Pppc4dmid}} (Release 4.0). We have then employed these ingredients to derive improved constraints based on current \PAMELA\ data (in sec.~\ref{sec:current_constraints}) and improved sensitivities for \AMS\ (in sec.~\ref{sec:future_sensitivity}). The constraints show that the subtle effects, ELDR in particular, modify the bounds by up to a factor of a few (typically loosening them) even at large masses. The sensitivity analysis shows that \AMS\ will be able to improve on current constraints by up to more than one order of magnitude and probe the thermal annihilation cross section for DM masses as large as 150 or 300 GeV (depending on the actual results). Perhaps more importantly, our analysis also shows that appropriately including the subdominant effects is crucial for a correct interpretation of the data: without them, one can e.g. easily obtain a `false positive' in favor of a DM signature. We have presented a concrete example where this indeed happens.

We are aware that we have only dealt here with a subset of the possible refinements for the calculation of the DM antiproton signal and its
background. Cosmic ray transport may also be a source of sizable uncertainty, which has been so far constrained from probes such as the B/C ratio. As the
production of antiprotons from high-energy protons and helium nuclei takes place in the same astrophysical sites as the fragmentation of carbon into
boron, using the B/C ratio has proved to be an efficient tool to constrain the antiproton predictions. We anticipate that the antiproton background will not
be much affected by the assumptions about the geometry of the magnetic halo. This is less clear for the antiproton DM signal. 
Alternatively, assuming that spallation reactions take place inside the acceleration sites and yield an astrophysical component of primary antiprotons could possibly
modify the DM predictions.

\medskip

In conclusion, our work shows the non-trivial complications connected to the use of antiprotons from astrophysics and from DM, especially at low DM masses, but also their very important probing power, if correctly mastered, for the upcoming future of DM indirect searches.

\bigskip

{\footnotesize
\paragraph{Acknowledgements}
We thank Marco Taoso for useful discussions and Timur Delahaye for providing us with the fit in~\cite{Timur}. MC and GG acknowledge the hospitality of the Institut d'Astrophysique de Paris, where part of this work was done.
\noindent Funding and research infrastructure acknowledgements: 
\begin{itemize}
\item[$\ast$] European Research Council ({\sc Erc}) under the EU Seventh Framework Programme (FP7/2007-2013)/{\sc Erc} Starting Grant (agreement n.\ 278234 --- `{\sc NewDark}' project) [work of MC and GG],
\item[$\ast$] French Institut universitaire de France (IUF) [work of MB and PS],
\item[$\ast$] French national research agency {\sc Anr} under contract {\sc Anr} 2010 {\sc Blanc} 041301 [work of MC].
 \end{itemize}
}

\bigskip

\footnotesize
\begin{multicols}{2}

\end{multicols}
\end{document}